\documentclass[a4paper,12pt]{spieman}  
\usepackage[utf8]{inputenc}   
\usepackage{soul} 
\usepackage{amsmath,amsfonts,amssymb}
\usepackage{aas_macros}
\usepackage{graphicx}
\usepackage{listings}
\usepackage{xcolor}
\usepackage{xspace}
\usepackage{setspace}
\usepackage{tocloft}
\usepackage{lineno}
\usepackage{paralist}
\usepackage[scaled=0.9]{helvet}
\usepackage[free-standing-units]{siunitx}

\renewcommand{\hl}{} 

\DeclareSIUnit\pixel{px}

\lstdefinestyle{pythonstyle}{
    language=Python,
    basicstyle=\ttfamily\footnotesize,
    keywordstyle=\color{blue},
    stringstyle=\color{red},
    commentstyle=\color{gray},
    numbers=left,
    numberstyle=\color{gray},
    stepnumber=1,
    breaklines=true,
    frame=single
}

\lstdefinestyle{xmlstyle}{
    language=XML,
    basicstyle=\ttfamily\footnotesize,
    keywordstyle=\color{blue},
    stringstyle=\color{red},
    commentstyle=\color{gray},
    numbers=left,
    numberstyle=\color{gray},
    stepnumber=1,
    breaklines=true,
    frame=single
  }

\title{SHINS, the SHARK-NIR Instrument Control Software}

\author[*,a]{Davide~Ricci}
\author[a]{Fulvio~Laudisio}
\author[a]{Alessandro~Lorenzetto}
\author[b]{Marco~De~Pascale}
\author[a]{Andrea~Baruffolo}
\author[a]{Daniele~Vassallo}
\author[a]{Domenico~Barbato}
\author[a]{Maria~Bergomi}
\author[c]{Florian~Briegel}
\author[a]{Elena~Carolo}
\author[a]{Simone~Di Filippo}
\author[a]{Marco~Dima}
\author[a]{Valentina~D'Orazi}
\author[a]{Tania Sofia~Gomes Machado}
\author[a]{Davide~Greggio}
\author[a]{Luca~Marafatto}
\author[a]{Dino~Mesa}
\author[c]{Lars~Mohr}
\author[a]{Gabriele~Rodeghiero}
\author[a]{Kalyan~Kumar~Radhakrishnan~Santhakumari}
\author[d]{Gabriele~Umbriaco}
\author[a]{Valentina~Viotto}
\author[a]{Jacopo~Farinato}

\affil[a]{INAF -- Osservatorio Astronomico di Padova, Vicolo
  dell'Osservatorio 5, I-35122, Padua, Italy }

\affil[b]{Leibniz Supercomputing Center of the BAdW Boltzmannstra\ss e 1,
  D-85748, Garching b. München, Germany}

\affil[c]{Max-Planck-Institut für Astronomie, Königstuhl 17,
D-69117 Heidelberg, Germany}

\affil[d]{Dipartimento di Fisica e Astronomia ``Augusto Righi'' --
  Alma Mater Studiorum Università di Bologna, via Piero Gobetti 93/2 -
  40129, Bologna, Italy}

\cftpagenumbersoff{figure}
\cftpagenumbersoff{table}

\begin{document}
\maketitle

\begin{abstract}

  SHARK-NIR is a new compact instrument for coronagraphic imaging,
  direct imaging, and coronagraphic spectroscopy in the near-infrared
  wavelengths ($0.96$--$1.7\um$) mounted at the left bent Gregorian
  focus of the Large Binocular Telescope (LBT).
  Taking advantage of the telescope's adaptive optics
  system\cite{2016SPIE.9909E..3VP}, it provides high contrast imaging
  with coronagraphic and spectroscopic capabilities and is focused on
  the direct imaging of exoplanets and circumstellar discs.

  We present SHINS, the SHARK-NIR instrument control software, mainly
  realized with the TwiceAsNice framework from MPIA - Heidelberg and
  the Internet Communications Engine (ICE) framework using the
  \texttt{C++} programming language.  We describe how we implemented
  the software components controlling several instrument subsystems,
  through the adaptation of already tested libraries from other
  instruments at LBT, such as LINC-NIRVANA.

  The scientific detector comes with its own readout electronic and
  control software interfaced with our software through
  Instrument-Neutral Distributed Interface (INDI)\footnote{\hl{Downey,
    E. C., ``INDI: Instrument-Neutral Distributed Interface.,'' (2007)}
    \url{http://www.clearskyinstitute.com/INDI/INDI.pdf}}.

  We describe the \texttt{C++} core software Observation Control
  Software, responsible for dispatching commands to the subsystems,
  also implementing a software solution to avoid a potential collision
  between motorized components, fully transparent to final users. It
  exposes an ICE interface and can be controlled by clients developed
  in different languages.

  Observation, calibration, and maintenance procedures are implemented
  by means of template scripts, written in \texttt{python} language,
  controlling Observation Control Software through its ICE
  interface. These templates and their parameters are configured using
  ``ESO-style'', \texttt{XML} Observation Blocks (OBs) prepared by
  observers, or in general SHARK-NIR users.

  The high-level control is carried out by REST HTTP APIs implemented
  in a \texttt{python} back-end, also acting as a web server for the
  several browser-based front-end Graphical User Interfaces, that
  allow the OBs to edit and sequence, as well as individual device
  movement and monitoring.

  Finally, we present the first scientific results obtained by
  SHARK-NIR using coronagraphic mode.

\end{abstract}


\section{Introduction}
\label{sec:introduction}

\begin{figure}[t]
  \centering
  \includegraphics[width=\textwidth]{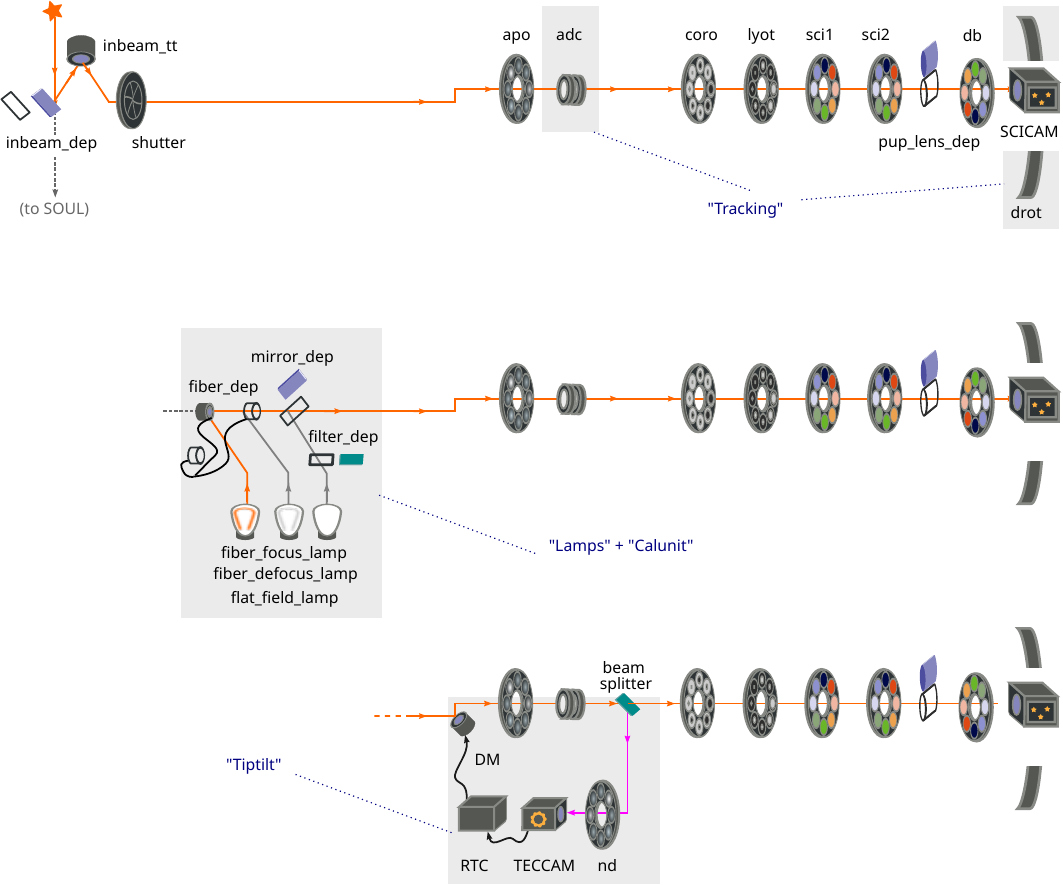}
  \caption{\label{fig:shark-opto} Schematic view of SHARK-NIR,
    \hl{replicated three times following the three principal ``light
      paths''.
      The top part shows the path of the light from a
      celestial source through until the science detector, and highlights
      the tracking devices.
      The middle part shows the the path from an internal calibration
      lamp through a fiber deployer, which are two of the components
      of the calibration unit.
      The bottom part is focused on the internal adaptive optics loop
      system of the instrument.
    }
    See Sect.~\ref{sec:introduction} for details. }
\end{figure}

SHARK-NIR is the near-infrared arm of SHARK, an instrument proposed
for the Large Binocular Telescope (LBT) in the framework of the ``2014
Call for Proposals for Instruments Upgrade and New Instruments'' and
composed of two channels (the second twin being called
SHARK-VIS\cite{2016SPIE.9908E..32P}), installed one for each LBT
side\cite{
   2016SPIE.9911E..27V, 
   2016SPIE.9909E..31F,  
   2015IJAsB..14..365F, 
   2018SPIE10701E..2BC 
}.
This compact instrument is mounted at the left bent Gregorian focus,
and exploits the Single conjugated adaptive Optics Upgrade for LBT
(SOUL)\cite{2016SPIE.9909E..3VP}, operating in a wavelength range
between $0.96$ and $1.7\um$ and focused on high contrast imaging
(HCI). It provides coronagraphic imaging, direct imaging, and
coronagraphic spectroscopic capabilities.
In this introduction, we describe its opto-mechanical components
following the path of the light through the instrument (see a
schematic view in Fig.~\ref{fig:shark-opto}). The description is
structured to reflect the hardware and software control that will be
expanded in the following sections.

\begin{description}

\item[Motion devices]  

  The light beam coming from the tertiary mirror of LBT \hl{(see top
    part of} Fig.~\ref{fig:shark-opto}) is split by a \texttt{PI
    LS-180} deployable dichroic (\textsf{inbeam\_dep}).  The visible
  wavelength proceeds to SOUL; the remaining part is reflected towards
  an ``entrance'' \texttt{PI M-232} tip-tilt mirror
  (\textsf{inbeam\_tt}) until it reaches the Maxon \texttt{GP16}
  instrument's shutter (\textsf{shutter}).
  Then, the beam passes through the coronagraphic system provided by a
  set of \texttt{PI M116 DGH} apodizers, coronagraphic mask, and Lyot
  stop (\textsf{apo}, \textsf{coro}, \textsf{lyot}), each set being
  mounted on a separate motorized wheel.
  These wheels also account for ancillary devices to carry out optical
  quality calibration procedures; slits for spectroscopy; and
  corresponding dispersing elements. \hl{This allows coronagraphic
    long slit spectroscopy at low and middle resolution, also
    involving a separate} \texttt{STANDA 8MT30-50DCE} pupil lens
  deployer \textsf{pup\_lens\_dep}.
  Afterwards, a set of \texttt{PI M-232} scientific wide,
  narrow-band, neutral density and dual-band filters are available on
  three separate wheels, (\textsf{sci1}, \textsf{sci2}, \textsf{db})
  allowing several filters combinations before reaching the Teledyne
  HAWAII H-2RG Scientific Camera (\textsf{SCICAM}).

\item[Tracking devices] 

  Field derotation (DROT) and Atmospheric Dispersion Correction (ADC)
  are ensured by tracking devices: a Fulling Motor \texttt{57SH76-4AM}
  (\textsf{drot}) and two combined \texttt{STANDA 8MPR16-1}
  (\textsf{adc}).
  The whole instrument is installed onto a large bearing, allowing its
  rotation around an axis passing through the center of the field of
  view, following an arbitrary trajectory. The derotation trajectory
  is obtained from the Telescope Control Software (TCS) and \hl{allows
    to follow and compensate the rotation of the sky}.
  The ADC, consisting of two counter-rotating prisms, is placed
  between the apodizer and the coronographic wheel. This correction is
  applied at a slower rate and allows the compensation of the
  atmospheric dispersion.

\item[Lamps and Calunit devices] 

  The instrument is equipped with a calibration unit \hl{(see the
    middle panel of} Fig.~\ref{fig:shark-opto}): one flat field lamp
  and two fiber lamps (one to provide an in-focus source and one for a
  defocus source) used for calibration (\textsf{flat\_field\_lamp},
  \textsf{fiber\_focus\_lamp}, \textsf{fiber\_defocus\_lamp}), and
  disposed in the optical path using a \texttt{PI M403.2DG} deployer
  (\textsf{fiber\_dep}) and two \texttt{PI M112-1.DG} deployers
  (\textsf{mirror\_dep}, \textsf{filter\_dep}).

\item[Tiptilt devices] 

  The instrument is \hl{also} equipped with an internal wavefront
  control system \hl{(see the bottom part of}
  Fig.~\ref{fig:shark-opto}): a beam splitter between the ADC doublet
  and the coronagraphic wheel \hl{directs a fraction of the beam
    through a} \texttt{PI M116 DGH} neutral density filter wheel
  (\textsf{nd}), and then to a FirstLight CRED-2 Technical Camera
  (\textsf{TECCAM}), which acts as a Wavefront Sensor for Real-Time
  Computer (\textsf{RTC}) provided by Microgate, calculating the
  real-time controls for a ALPAO 97-15 Deformable Mirror
  (\textsf{DM}).
  This system ensures fast tip-tilt correction \hl{up to $2\kHz$}, Non
  Common Path Aberrations (NCPA) correction, and fine Point Spread
  Function (PSF) positioning behind the coronagraph.

\item[SCICAM]  

  The scientific detector is an Hawaii 2RG $2048 \times 2048 \pixel$
  from Teledyne with $18\um$ pixel size, with a
  $18\times18^{\prime\prime}$ field of view and a pixel scale of
  $14.5$ milliarcseconds per pixel. The detector operates under
  cryogenic temperature and is therefore contained inside a
  cryostat. Temperature is controlled by a Lake Shore 336.

\end{description}


The main scientific goal of SHARK-NIR is the imaging (direct,
coronagraphic) and long slit spectroscopy of exoplanets, their
detection, and characterization. Other science objectives include
brown dwarfs, protoplanetary discs, stellar jets, quasi-stellar
objects and active galactic nuclei.
SHARK-NIR started its scientific operations in 2024, after a wider
assembly, integration, and test phase\cite{
  2024SPIE13096E..3TC, 
  2024SPIE13096E..1WB, 
  2022SPIE12187E..09B, 
  2022SPIE12185E..8IV, 
  2022SPIE12185E..6WU, 
  2022SPIE12185E..22F, 
  2022SPIE12184E..3VM  
}
and the commissioning, soon obtaining promising scientific results.
These scientific results\cite{2025A&A...693A..81B,
  2025MNRAS.536.1455M} are the contribution to the characterization of
an eccentric giant planet detected through radial velocity (RV) around
the star \textsf{HD~57625}, and a deep imaging study of three
accelerating stars: \textsf{HIP~11696}, \textsf{HIP~47110} and
\textsf{HIP~36277}.


In this paper, we present in detail SHINS, the SHark INstrument
control Software, expanding and structuring the improvements described
in a dedicated series of SPIE proceedings\cite{
  lorenzetto2024, 
  2022SPIE12189E..20R, 
  2020SPIE11452E..1TD, 
  2018SPIE10707E..1MD 
}.

\section{Control Electronics}
\label{sec:control-electronics}

\begin{table}[b]
    \centering
    \begin{tabular}{lll}
\hline
             \textbf{Device}              &      \textbf{Description}       & \textbf{Motor/control} \\
\hline
         \textbf{Motion devices}          &                                 & \\
          \textsf{inbeam\_dep}            &       Deployable dichroic       & \texttt{PI LS-180}  \\
           \textsf{inbeam\_tt}            &    Entrance tip-tilt mirror     & \texttt{PI M-232}\\
            \textsf{shutter}              &       Instrument shutter        & \texttt{Maxon GP16} \\
              \textsf{apo}                &         Apodizer wheel          & \texttt{PI M-116 DGH} \\
              \textsf{coro}               &    Coronagraphic mask wheel     & \texttt{PI M-116 DGH} \\
              \textsf{lyot}               &    Lyot stop and grism wheel    & \texttt{PI M-116 DGH} \\
         \textsf{pup\_lens\_dep}          &       Pupil lens deployer       & \texttt{STANDA 8MT30-50DCE} \\
\textsf{sci1}, \textsf{sci2}, \textsf{db} &    Scientific filter wheels     & \texttt{PI M116 DGH} \\
\hline
        \textbf{Tracking devices}         &                                 & \\
              \textsf{drot}               &         Field derotator         & Fulling Motor \texttt{57SH76-4AM} \\
              \textsf{adc}                &           ADC prisms            & \texttt{STANDA 8MPR16-1} \\
\hline
   \textbf{Lamps and Calunit devices}     &                                 & \\
       \textsf{flat\_field\_lamp}         &   Flat field calibration lamp   & (CyberPower PDU) \\
       \textsf{fiber\_focus\_lamp}        & Fiber lamp for in-focus calib.  & (CyberPower PDU) \\
      \textsf{fiber\_defocus\_lamp}       & Fiber lamp for defocused calib. & (CyberPower PDU) \\
          \textsf{fiber\_dep},            &      Optical path deployer      & \texttt{PI M-403.2DG} \\
          \textsf{mirror\_dep},           &      Optical path deployer      & \texttt{PI M-112-1.DG} \\
          \textsf{filter\_dep}            &      Optical path deployer      & \texttt{PI M-112-1.DG} \\
\hline
        \textbf{Tiptilt devices}          &                                 & \\
               \textsf{nd}                &  Neutral density filter wheel   & \texttt{PI M-116 DGH} \\
             \textsf{TECCAM}              &    Wavefront sensing camera     & \texttt{FirstLight C-RED2} \\
              \textsf{RTC}                &     RTC Basic Control Unit      & \texttt{Microgate}  \\
               \textsf{DM}                &   DM for tip-tilt correction    & \texttt{ALPAO 97-15} \\
\hline
             \textsf{SCICAM}              &       Scientific detector       & \texttt{Teledyne HAWAII H-2RG}  \\
\hline
\end{tabular}
\caption{\hl{SHARK-NIR device descriptions, and related
    hardware. Calibration lamps are switched on and off trhough a
    CyberPower Power Distribution Unit. See}
  Sect.~\ref{sec:introduction} and~\ref{sec:control-electronics} for
  details.}
    \label{tab:devices}
\end{table}

In this section, we present the hardware responsible for controlling
the optomechanical parts, the lamps, the wavefront control system and
the cameras of the instrument. \hl{See Table}~\ref{tab:devices}
\hl{for hardware details}, and the gray squared boxes in
Fig.~\ref{fig:shark-network}).

\subsection{MoCon boards}

All motorized functions (\emph{Motion devices}, \emph{Calunit
  devices}, and \emph{Tracking devices}) are electronically controlled
by a group of three MoCon boards, custom-made and provided by the
Max-Planck-Institut für Astronomie (MPIA) in Heidelberg, which are the
same type motion control boards already employed by the
LINC-NIRVANA\cite{2008SPIE.7013E..26H} instrument at LBT.

\hl{The three MoCon boards can be operated from outside using a
  \texttt{telnet} connection and a list of predefined commands. They
  can mount modules to drive dc motors in closed loop or stepper
  motors in open loop. Can read feedback from incremental encoders and
  absolute encoders.  Each board can control up to 8 modules, each
  with up to three sensors, two ``end of stroke'' and a home
  sensor. The sensors can be ``normally open'' switches, ``normally
  closed'', or level switches. The firmware of the MoCon automatically
  reads the incremental encoders and corrects the power using a
  Proportional-Integral-Derivative (PID) controller.}
The first board is dedicated to the \emph{Motion devices} wheels and
the shutter: the second is dedicated to the \emph{Motion devices}
``entrance'' tip-tilt mirror and deployers, as well as the
\emph{Calunit devices} deployers; the third and last is dedicated to
the \emph{Tracking devices} derotator and ADCs.

\hl{The instrument has 7 filter wheels, 6 linear stages, one tip tilt
  actuator, moreover there is the ADC with two motorized prisms and
  the derotator} (see Table~\ref{tab:devices} for details).
\hl{The
  wheels have one reference sensor used as home position while the
  linear stages have two end of stroke switches, while the ADC prisms
  are mounted on circular stages, with stepper motors equipped with a
  reference sensor.  The shutter is the only linear stage built from
  scratch using a DC motor connected to an encoder and to a threaded
  rod through a reduction gearbox. Two induction sensors act as end of
  strokes. It is planned to be refurbished with a commercial shutter.

   The derotator is slightly more complex, using a custom bearing
  connected in parallel, in order to minimize backlash, to two high
  torque stepper motors. Two induction sensors act as end of stroke,
  and an absolute encoder with $2^{16}$ positions is connected to the
  bearing by means of a zero backlash gearbox. Each line of the
  absolute encoder corresponds to $\approx 20\arcsecond$. A special
  procedure to minimize backlash is performed at the beginning of the
  observations. This procedure required a customization of one of the
  MoCon boards that allowed moving the two stepper motors, controlled
  by a single module, independently in the opposite directions to
  ensure a tight fit between the motor gears and the main bearing.}

\subsection{Lamps, Adaptive optics, and scientific camera}


\emph{Lamps} are switched on and off by an off-the-shelf Power
Distribution Unit (PDU) produced by the CyberPower company.  We also
plugged to the same PDU: the three MoCon boards, the scientific camera
electronics, the DM electronics the RTC, the TECCAM and its related
components and the Lake Shore controller, so that it was possible to
set up a shutdown/startup script for maintenance purposes, as well as
power cycle individual components as troubleshooting procedure.


The internal wavefront control system of the \emph{tiptilt devices}
include the DM, the TECCAM, and the RTC. The DM is a 97 actuators DM
from ALPAO which comes with its own control electronics. The
FirstLight CRED-2 TECCAM, \hl{that has been set to operate at a
  temperature of $-40\degree$, runs at a frame rate up to $400 \Hz$ in
  full frame and up to $2 \kHz$ depending on the subframe's
  dimension. The choice of the subrframe depends wether the camera is
  used for tiptilt correction or in ``patrol'' mode, and it is
  detailed in } Sect.~\ref{sec:rtc-panel}. The RTC, called Basic
Control Unit (BCU) and developed by Microgate, has a custom
motherboard with a Camera Link port, that receives the frames from the
TECCAM. The BCU motherboard also has a Molex connector compliant with
the DM control electronics. A dedicated firmware detects the light
centroid variation in the frames from the technical camera and applies
the proper correction to the deformable mirror.


Finally, the SCICAM is controlled by a Multi-Purpose Control and
Interface Electronics (MACIE) by Markury Scientific Inc.~connected to
the detector workstation (\texttt{sashaws}) where runs the control
software and the INDI server.

\begin{figure}[b]
  \centering
  \includegraphics[width=\textwidth]{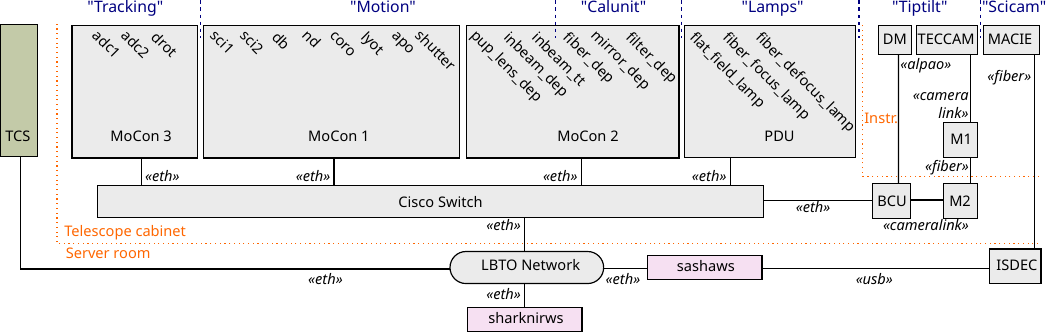}
  \caption{\label{fig:shark-network} How the SHARK-NIR devices \hl{(gray
    boxes), workstations (pink boxes), and TCS (olive green box),
    described} in Sect.~\ref{sec:introduction}, are controlled by
    Control Electronics and interfaced in the Control Network to the
    Instrument workstation. The location of the hardware is also
    specified in clear between dotted lines. See
    Sect.~\ref{sec:control-electronics}-\ref{sec:control-network} for
    details. }
\end{figure}

\section{Control Network}
\label{sec:control-network}

In this section, we describe how the control electronics are embedded
in the control network of the instrument (see
Fig.~\ref{fig:shark-network} for details).


The three MoCon boards, which are located in a cabinet close to the
instrument, are Ethernet-connected to the SHARK-NIR Cisco Switch, that
connects to the LBTO (LBT Observatory) internal network via standard
Ethernet LAN.
The PDU controlling the lamps is also connected to the same Switch.
In the same cabinet are placed the control electronics running the
RTC: the communication between the DM and the BCU takes place on ALPAO
dedicated cable, while the communication between the TECCAM and the
BCU are sent via CameraLink and mediated by a Media Converter (M1 and
M2 in Fig.~\ref{fig:shark-network}) via optical fiber.  Then, the BCU
communicates via Ethernet through the Switch.


The scientific camera is controlled by a dedicated component, run by a
separate workstation, and developed by the Steward Observatory in
Tucson.  Direct communication between camera electronics (Teledyne
SIDECAR) and its workstation is realized using a dedicated FPGA-based
interface, called ISDEC, via optical fiber on camera side, and via USB
on workstation side (see Fig.~\ref{fig:shark-network}).  This
workstation is called \texttt{sashaws} and it is placed in the LBTO
server room.


Then, the control software of SHARK-NIR is managed by an Instrument
workstation located in the same server room.  Control information are
transmitted through the LBTO network node, via standard Ethernet LAN,
to the controllers connected to the Switch and the scientific camera
workstation.


Fig.~\ref{fig:shark-network} also shows the \hl{TCS} workstation,
which is not part of SHARK-NIR, \hl{but it is used as a resource by
  the instrument control software}, as described in
Sect.~\ref{sec:subsyst-contr-softw}.

\hl{SHARK-NIR network devices belong to a dedicated VLAN. From the
  LBTO control network, it is possible to directly connect to the
  instrument devices; while from outside the control network a VPN
  access is needed.  Most of SHARK-NIR network devices are protected
  by telnet login (Mocon boards) and HTTP BasicAuth credentials (PDU,
  CISCO Switch, and other web interfaces), which is enough, once in
  the local network, to avoid mistakes while operating other
  instruments.}

\section{Subsystems Control Software}
\label{sec:subsyst-contr-softw}

\begin{figure}[t]
  \centering
  \includegraphics[width=\textwidth]{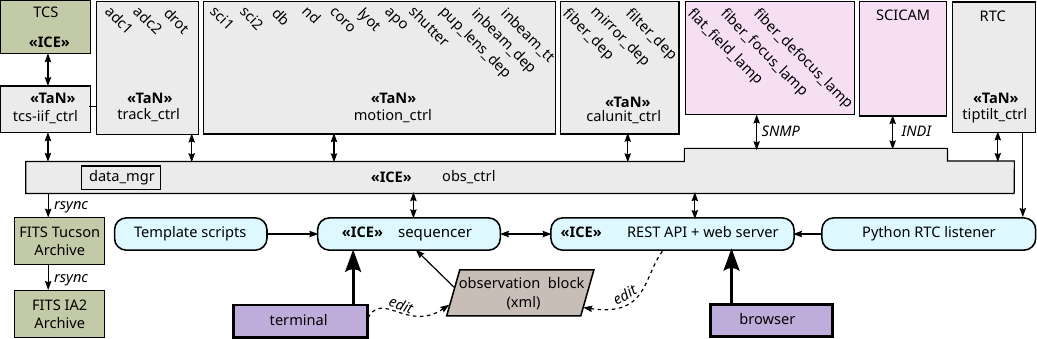}
  \caption{ \label{fig:shark-stack} SHARK-NIR software stack. See
    Sect.~\ref{sec:subsyst-contr-softw},
    ~\ref{sec:observ-contr-softw}, ~\ref{sec:sequ-templ-observ},
    ~\ref{sec:rest-api}, ~\ref{sec:web-based-graphical}, for
    details. In the top part, gray boxes refer to \texttt{C++}-based
    control developed over TaN and ICE frameworks, \hl{while pink
      boxes are for non-C++ code interfaces, and their external
      Communication protocols are shown out of the boxes. TCS, FITS
      Tucson Archive, and FITS IA2 Archive, shown in olive green boxes
      on the left, refer to external interfaces.}  Rounded boxes refer
    to \texttt{python} code, some of which use ICE to connect to the
    central \texttt{C++} component \texttt{obs\_ctrl}. Configuration
    files for Templates are also shown in the slanted box.  Finally,
    \hl{bold purple boxes} show two different ways to control SHINS:
    via terminal commands, manually editing custom Templates setup for
    OB creation and then launching the sequencer; or via the web
    browser. Browser also allows the XML Observation Blocks editing
    and the control of individual components, as well as and the
    monitoring of the RTC diagnostic data via a dedicated
    \texttt{python} listener. }
\end{figure}

\hl{LBT allows a high degree of flexibility to the instrument teams
developing the architecture of the control software for their
instruments, although they do express a preference for web-based user
interfaces.}  Given the experience of our team with ESO VLT control
software, we organized the architecture of SHINS in a way similar to
that of ESO VLT instruments: dedicated \emph{Subsystems Control
  Software} refer to a central component, called \emph{Observation
  Control Software}, which is in charge of coordinate them and
dispatching commands.  On top of this component, observation,
calibration, and maintenance procedures are implemented by means of
\emph{Template} scripts, which are called by a \emph{sequencer}
through a set of parameters.

We now describe in detail the control software of the several
subsystems. Fig.~\ref{fig:shark-stack} shows a schematic view of the
control software.

\begin{description}

\item[\texttt{motion\_ctrl}] \emph{service} controls \emph{Motion
    devices}.
  As we used MoCon boards to control motors, we also decided to adopt
  its high-level control framework, also developed by MPIA for the
  LINC-NIRVANA instrument, called TwiceAsNice (TaN) in the form of a
  set of \texttt{C++} \hl{libraries} \cite{2008SPIE.7019E..1TB}.
  A common feature of all the sub-packages developed using the TaN
  framework is that each sub-package is mapped to a TaN \emph{service}
  available through the TaN server process, each service controlling a
  subset of SHARK-NIR devices.  \hl{The adoption and customization of
    TaN libraries represent the most of the code reuse of SHINS}.

  \hl{The \emph{service} loads the configuration for each motor,
    opens the \texttt{telnet} communication with the MoCon boards and
    keeps track of the status of the motor, the position, the speed,
    the status of the switches and the errors. In the configuration
    for each motor can be specified the ranges of the motor (software
    limits), the named positions, different units to control the
    device and the rules to convert these units.}

  \texttt{motion\_ctrl} is then the service dedicated to \emph{Motion
    devices}, the devices used by the instrument during science
  operations.

\item[\texttt{calunit\_ctrl}] is the \emph{service} controlling the
  motors used during calibration operations, i.e. the \emph{Calunit
    devices}.

\item[\texttt{track\_ctrl}] is the \emph{service} controlling the
  derotator and the two ADCs, i.e. the \emph{Tracking devices}.
  However, in this case, there is the need to calculate the position
  information based on the telescope's position at a given time.
  Tracking functions follow the telescope movement, thus are active
  during observation, and controlled by the \texttt{track\_ctrl}
  component. SHINS implements two tracking functions: field derotation
  and atmospheric dispersion correction.

  The derotator is one of the most critical devices, since it has to
  remain as much in sync with the telescope as possible; for this
  reason the trajectory is queried and uploaded to the derotator
  \hl{every second.}
  \hl{The derotator \emph{service} addresses the peculiarity of the
    chosen setup. The bearing compensates for the apparent field of
    view rotation due the Alt-Az nature of the telescope.  Each of the
    8 MoCon channels can follow an arbitrary trajectory profile loaded
    in a shared memory on the board, so once the derotator
    \emph{service} obtains from the telescope interface the derotation
    profile (that depends on the current preset), it converts it in a
    series of cubic segments. These segments are sent to the MoCon
    firmware to approximate an arbitrary profile, keeping track of the
    position by reading the absolute encoder.}

  \hl{In contrast}, the correction for atmospheric dispersion,
  performed by the ADC, in SHARK-NIR is less critical than the
  derotator, and it does not need knowledge of the telescope
  trajectory. The ADC position is updated with a time scale of few
  seconds, and is a function of pressure, temperature, and zenith
  angle.

  For all these reasons, derotator trajectory and ADC position are a
  function of the telescope position, velocity, and acceleration.
  The LINC-NIRVANA control software team also expanded for SHARK-NIR
  the TaN service used as interface to the TCS, which was already
  developed for LINC-NIRVANA. This component is called
  \texttt{tcs\_ctrl} and it is described here below.

\item[\texttt{tcs\_ctrl}] is the \emph{service} interfacing with, and
  controlling the TCS.
  The Large Binocular Telescope provides a set of API for instruments
  control software to interface to TCS, called Instrument InterFace
  (IIF). Such APIs are declared in specification language for ICE
  (\texttt{SLICE}) and implemented in \texttt{C++}. The IIF APIs are
  accessible to SHINS thanks to a TaN \emph{service},
  \texttt{tcs\_ctrl}, developed and tested by MPIA for LINC-NIRVANA.

\item[\texttt{tiptilt\_ctrl}] is the \emph{service} controlling the
  internal wavefront control system of SHARK-NIR.
  The RTC is responsible for updating the tiptilt of the DM in
  real-time, so to reduce telescope vibrations. Moreover, in order to
  correct NCPA, it is possible to upload a static shape to the DM in
  the form of a vector of coefficient of Zernike polynomials, kept
  during the tip/tilt loop.  Moreover, it is also used for
  fine-positioning of the scientific object behind the coronagraphic
  mask. All of the communication happens through the BCU, and there is
  no direct communication with DM electronics or TECCAM.  The private
  company that produced the BCU, Microgate, provided control software
  made by a set of Matlab functions. These functions have been ported
  to \texttt{C++} and implemented in the \texttt{tiptilt\_ctrl}
  service logic so that it is possible to write directly in the BCU
  memory the required quantities, and send them from the workstation
  to the BCU with User Datagram Protocol (UDP) packets using the
  Microgate Protocol.
  At lower level, we reused the TaN libraries of
  ARGOS\cite{2019A&A...621A...4R}, the Binocular laser guided
  ground-layer adaptive optics at LBT, to communicate with the BCU,
  which prepares the UDP packets and sends them to BCU.

\end{description}

\section{Observation Control Software}
\label{sec:observ-contr-softw}

The TaN framework is built on top of Internet Communication Engine
(ICE), an object-oriented Remote Procedure Call (RPC)
framework\footnote{\url{https://zeroc.com/ice}}.  For this
reason, we developed in \texttt{C++} the central component of the
software, called \texttt{obs\_ctrl} or \emph{Observation Control
  Software}, on top of ICE.

\emph{Services} \hl{run in separate threads, allowing parallel
  operations (for example the instrument setup while the telescope is
  moving). Moreover, The ICE framework allows exposing both
  synchronous and asynchronous versions of all class methods of
  \texttt{obs\_ctrl}. This feature is highly used for high-level
  control} (see Sect.~\ref{sec:sequ-templ-observ}), ensuring
  non-blocking operations.

\texttt{obs\_ctrl} connects at startup to all other \emph{services}
and to the INDI server, and controls all the operations of
SHARK-NIR. The simultaneous communication with the subsystem services,
along with its direct interaction with the scientific detector allows
\texttt{obs\_ctrl} to control and execute all requested operations.

This component stores the state of each SHINS device \hl{in a
  dedicated object, which is updated by querying the PDU (via SNMP),
  and the \emph{services}, at a rate of $1\Hz$, while INDI pub/sub
  protocol provide the update of the state of the SCICAM based on
  events. So, it is possible for client components to retrieve the
instrument status without querying individual subsystems and to
  monitor state changes if the devices are operated externally. This
  system provides an instant status for SHINS, while an external alert
system, based on
EPICS}\footnote{\url{https://pyepics.github.io/pyepics/overview.html}},
\hl{is feeded by \texttt{cron} jobs with telemetry abotut temperature and
pressure of the cryogenic system and is monitor by LBTO}.

Apart coordinating and interfacing with the \emph{services} subsystems
described in the previous section, the Observation Control Software
implements embedded controls and checks that are discussed in the
following subsections.

\subsection{SCICAM and PDU control}
\label{sec:scicam-pdu-control}

The scientific camera control software is developed directly in
\texttt{obs\_ctrl} using the Instrument Neutral Distributed Interface
 pub-sub protocol, already used at LBTO to control the LBTI
scientific camera, LMIRCam\cite{2012SPIE.8446E..4FL}.

Lamps are switched on and off by controlling their PDU through Simple
Network Management Protocol (SNMP). Two separate \texttt{obs\_ctrl}
methods provide setting and getting the status of the lamps. Given the
simplicity of the device, no separate \emph{service} has been set up
to control these components.

\subsection{Anti-collision software}
\label{sec:anti-coll-softw}

During the AIT phase, a potential collision issue between two
motorized components emerged, namely the coronagraphic motor wheel and
the fiber deployer. We developed a software solution at the
\texttt{obs\_ctrl} level, that is also available while using other
software such as engineering panels. The anti-collision system is
based on three protection layers:

\begin{enumerate}

\item The first level of protection consists in the fact that the
  wheel before making any movement checks the position of the fiber
  deployer, and if this is below a certain threshold it raises an
  exception.

\item The second level of protection consists of a couple of functions
  that intervene during the setup phase of the instrument. The setup
  is classified according to wheel and deployer movements, and is
  divided into nine scenarios. Of these nine scenarios, five do not
  need any special attention, while for the last four, the software
  splits the setup into two or three consecutive steps where the
  coronagraphic wheel never moves when the fiber is in the cut-off
  region, minimizing execution times. Eventually, the fiber is moved
  to the safe region and then returned to the initial position at the
  end of the wheel setup.

\item The third level of protection is based on a set of callbacks
  installed on the services controlling the motors. Whenever the
  coronagraphic slit wheel starts any movement the framework calls
  back the Observation Software which in turn launches a new thread
  that controls continuously the calibration fiber deployer
  position. If it is beyond the interdiction threshold, it commands an
  abort the movement and move the fiber deployer into the safe
  area. This third level of protection intervenes whenever a movement
  of the coronagraphic wheel is commanded by \textit{any} client of
  the \emph{services} provided that SHINS is up and running, thus
  protecting the instrument also from accidental errors of an operator
  working with the engineering GUIs.

\end{enumerate}
The anti-collision system is driven by the Observation Control
Software, and relies on the position of the calibration fiber
deployer; for this reason at the startup of the Observation Software
such deployer is always homed.

\subsection{NCPA Rotation}
\label{sec:ncpa-rotation}

Each SHARK-NIR observation estimates and corrects the NCPA.  Due to
the structure of the instrument, it is necessary to distinguish
between two types of NCPA: internal-\emph{static} NCPA and
external-\emph{rotating} NCPA.

All SHARK-NIR devices excepted to \textsf{inbeam\_dep} and
\textsf{inbeam\_tt} are mounted on a rotating bearing, to allow
field-stabilized observations.  Therefore, they are integral to the
field rotation. From such path is retrieved the internal-\emph{static}
NCPA.
On the other hand, the portion of the path that ranges from the
telescope primary mirror to the instrument shutter is not involved in
the bearing rotation and it doesn't follow the field rotation. These
are called the external-\emph{rotating} NCPA, and result in an
aberration pattern that depends on the derotator angle.

These two complementary effects are isolated, modeled, and corrected.

Internal-\emph{static} NCPA, which are integral with the rotation and
are computed with the internal light source, are loaded on the DM as a
static shape, retrieved after an focal-plane wavefront sensing
technique called Phase Diversity\cite{2018SPIE10705E..16V}. By
applying a static shape, these NCPA are corrected.

External-\emph{rotating} NCPA are corrected by uploading other modes
on the DM, on top of the static shape. These modes are retrieved by an
optimization procedure based on a ``trial and error'' approach,
performed on every observation's target, and saved as a different
shape \hl{in the form of a vector of positions}.
This shape is software-rotated \hl{every second} based on the position
of the bearing (i.e. of the whole instrument), and suddenly \hl{sent
  to the BCU through a \texttt{tiptilt\_ctrl} command, which adds it
  to the calculated tip-tilt DM shape it creates internally}, to
compensate for the derotation effect and fine-optimize the optical
quality.

The whole DM shape is then maintained as a superposition of
internal-\emph{static} and external-\emph{rotating} NCPA. The RTC
architecture reserves the first two modes for fast tip-tilt loop,
therefore once the DM shape is defined, it is maintained even when the
internal tip-tilt loop is operating.

\hl{In the current ``early science'' phase, internal-\emph{static} and
  external-\emph{rotating} NCPA are calculated before each
  observation, in order to have a consistent statistics about the
  conditions of temperature and telescope position in which they are
  modeled. Based on this information, it is foreseen to set up the
  best strategy to minimize the time loss due to this optical quality
  calibration procedure.}

\subsection{The data Manager}
\label{sec:data-manager}

The Data manager, called \texttt{data\_mgr}, is a class of
\texttt{obs\_ctrl} responsible for scheduling the acquisitions of the
SCICAM, and retrieving the data and creating FITS file, using
\texttt{cfitsio}, reporting on the FITS header all the information of
the instrument status.  For what concerns header merging, the FITS
file for the observation produced by the scientific camera is received
by \texttt{obs\_ctrl} through the INDI interface. This is implemented
by listening to INDI TCP port, receiving the file as a ``blob'' of raw
bytes and writing them to disk; in this process is not possible to add
new FITS keywords to the header, thus a second blob is created by
\texttt{obs\_ctrl}, with no data and with a header populated by all
the required FITS keywords. The \texttt{data\_mgr} component receives
this last file and the FITS file produced by the scientific camera.
Moreover, this component merges the scientific camera header in the
full header produced by \texttt{obs\_ctrl} as well as copying the
data, using \texttt{cfitsio} library APIs.  Data are then copied to
the FITS Archive system in Tucson via \texttt{rsync} through a Network
File System (NFS) remotely-mounted directory, and from there to the
Italian center for Astronomical Archive (IA2) archive in
Trieste\footnote{\url{https://www.ia2.inaf.it/}}, Italy. Dedicated
\texttt{cron} jobs on the instrument workstation \texttt{sharknirws}
and on the scientific detector workstation \texttt{sashaws} provide a
rotation of the files to avoid disk saturation, on a timespan period
of 1 year.

\section{Sequencer, Templates, and Observation Blocks}
\label{sec:sequ-templ-observ}

\begin{figure}[p]

\begin{lstlisting}[style=xmlstyle]
<?xml version="1.0"?>
<ObservationBlock       xsi:schemaLocation="ObservationBlock.xsd"
    OBID="TestID"       ProgramID="TestProgID"
    PIName="El Condor"  PIID="TestPIID">

    <Template>
    <TPLID>TestForJatis</TPLID>
    <TPLName>Test for JATIS</TPLName>

    <InstrumentSetup>
      <SCI_FILT_W1>HOLE</SCI_FILT_W1>
      <SCI_FILT_W2>BB_H</SCI_FILT_W2>
    </InstrumentSetup>

    <DetectorSetup>
      <DIT>1</DIT>
      <NDIT>1</NDIT>
    </DetectorSetup>

    </Template>

    <Template>
    ... an additional template in the OB ...
    </Template>

</ObservationBlock>
\end{lstlisting}

\begin{lstlisting}[style=pythonstyle, basicstyle=\ttfamily\footnotesize]
from shinsapp import SHINSApp
from util import exception
from util.shinsLogger import logger

class TestForJatis(SHINSApp):

    @exception.handler_for_run
    def run(self, ins_s=None, det_s=None, tel_s=None, rtc_s=None):
        logger.info(__name__)
        obsController = self.obsController

        # Async instrument and detector setup
        async_ins = obsController.begin_setupInstrument(ins_s)
        async_det = obsController.begin_sashaSetup(det_s)

        # Waiting Instrument and detector setup
        self.status = obsController.end_setupInstrument(async_ins)
        self.status = obsController.end_sashaSetup(async_det)

        logger.info("Instrument and detector setup... completed")

        # Sync exposure
        self.status, self.filenames = obsController.sashaExpose()
        logger.info("FITS files are "+str(self.filenames))

\end{lstlisting}
\caption{\label{fig:code} Examples of OB in \texttt{XML} format (top),
  setting parameters for the corresponding template script in
  \texttt{Python} language (bottom). }

\end{figure}

SHARK-NIR is operated using Observing Blocks (OBs), the same way this
is done with instruments hosted in modern astronomical observatories
such as the VLT.

The ICE Framework allows to expose as \texttt{python} modules the
functions of \texttt{obs\_ctrl}. Then, on a higher layer, we developed
a sequencer (\texttt{seq}), a \texttt{phyton}-based application that
acts as a client to \texttt{obs\_ctrl}, and allows the execution of
the OBs.

OBs are logical blocks, stored as XML files; the content specifies the
sequence for automated operations, called \emph{Templates}, and the
parameters to be passed to them. Each template implements an
observation, calibration, or maintenance procedure. It is composed of
a reference file, that includes a short description of the template,
and all the parameters that are fixed during
the execution of the template; a signature file, describing all the
parameters the user can specify; and the template script itself, which
is the actual implementation of the procedure (see rounded boxes in
Fig.~\ref{fig:shark-stack}).

Templates scripts are implemented as \texttt{python} scripts accepting
five input arguments in the form of \texttt{python} dictionary, an
unordered list of labels (representing the parameters) and
values. Each dictionary specifies the setups as follows: Instrument
setup; Detector setup; Telescope setup; Real-time tip/tilt subsystem
setup; Observation/Program Info setup.

When an OB is loaded, \texttt{seq} process operates as follows: reads
the specified XML files and extract an ordered list of templates to
execute; for each template, the corresponding list of parameters is
then checked against its scheme. The resulting \texttt{python}
dictionaries are merged with the ones containing the fixed parameters
to obtain a complete setup.  Finally, the OB is executed.

\hl{In general, templates start from an initial instrument setup,
  followed by one or more science or technical images, that at each
  loop or phase of the script performs checks, moves the instrument to a
  different setup, or alternates a set of operations.
  Error handling during the execution of the template are managed at
  template and sequencer level by separate \texttt{python}
  decorators.

  At lower level, if an error occurs, such as the reaching of a limit
  switch, or an unavailability of a \emph{service}, the error is
  propagated to the sequencer and shown to the final user.

  At higher level, the status is checked at any relevant operation. As
  templates are complex operations involving the whole instrument and
  the telescope, the template is aborted if an error occurs and no
  recovery or error clearence is foreseen in case of failure: a
  talking error message is displayed in the template log so that the
  operator can verify it, and relaunch the OB.}

During execution, \texttt{seq} calls the list of \texttt{python}
scripts in the order specified by the OB, passing the corresponding
dictionary objects as a parameter. Fig.~\ref{fig:code} \hl{shows an
  example of OB calling a single template script.}

\section{REST API}
\label{sec:rest-api}

\begin{figure}[p]
  \centering
  \includegraphics[width=0.77\textwidth]{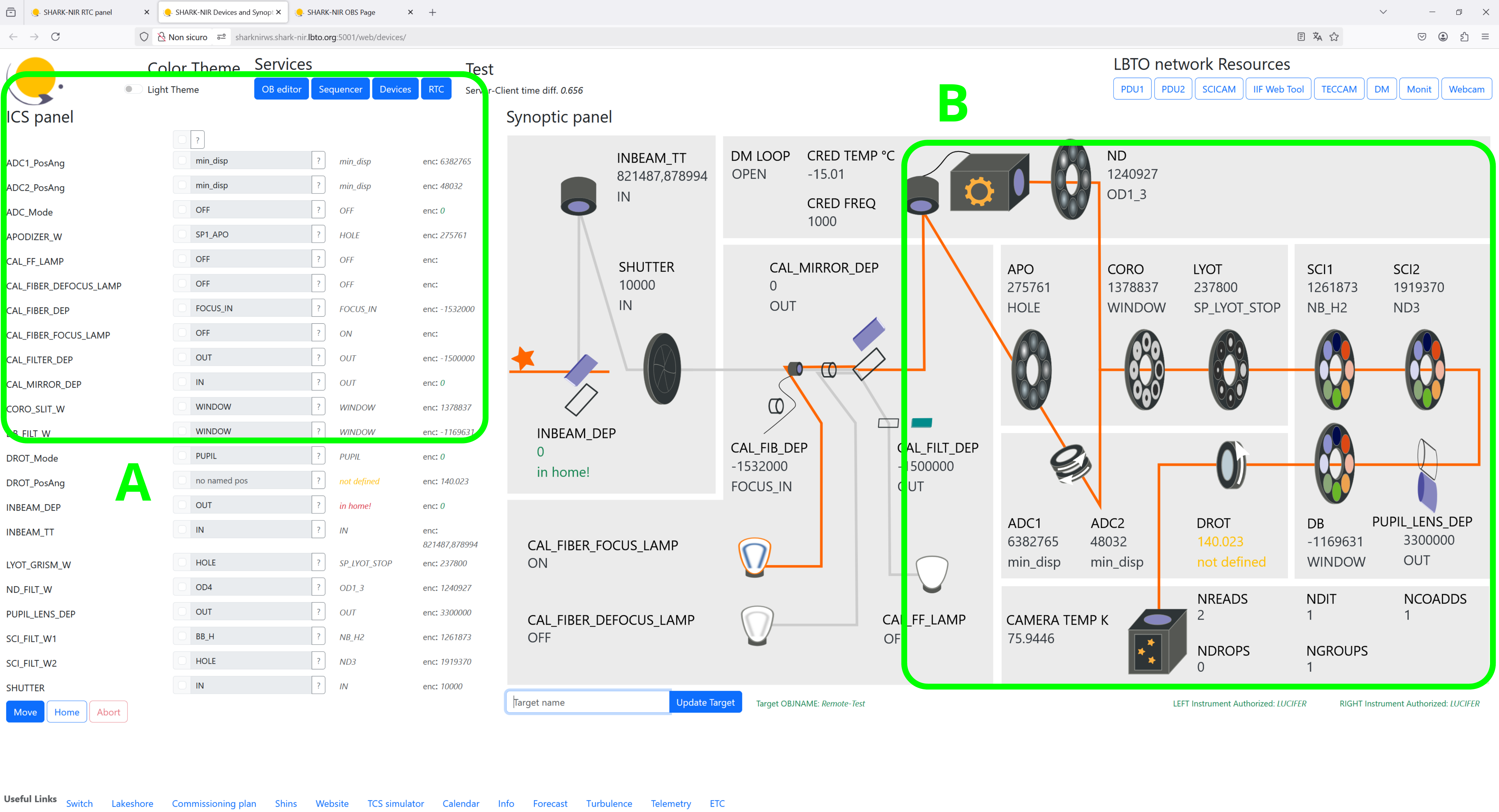}
  \caption{\label{fig:dev} Devices and Synoptic panel overview. See
    ``A'' (Fig.~\ref{fig:devices}) and
    ``B'' (Fig.~\ref{fig:synoptic})
    for details. }
  \includegraphics[width=0.77\textwidth]{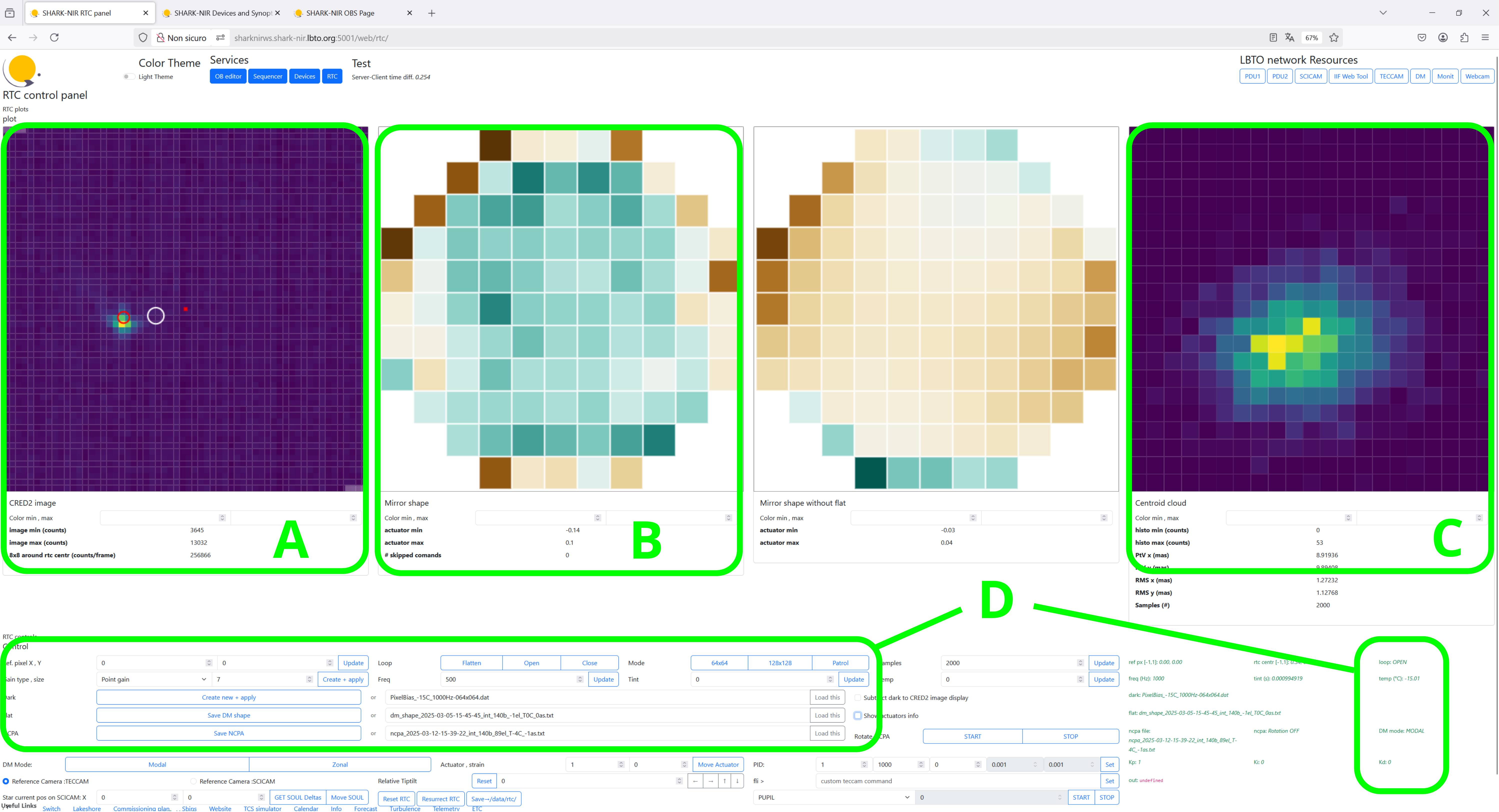}
  \caption{\label{fig:rtc} RTC Engineering panel overview. See
    ``A'' (Fig.~\ref{fig:rtc-cred}),
    ``B'' (Fig.~\ref{fig:rtc-dm}),
    ``C'' (Fig.~\ref{fig:rtc-cloud}), and
    ``D'' (Fig.~\ref{fig:rtc-control})
    for details.  }
  \includegraphics[width=0.77\textwidth]{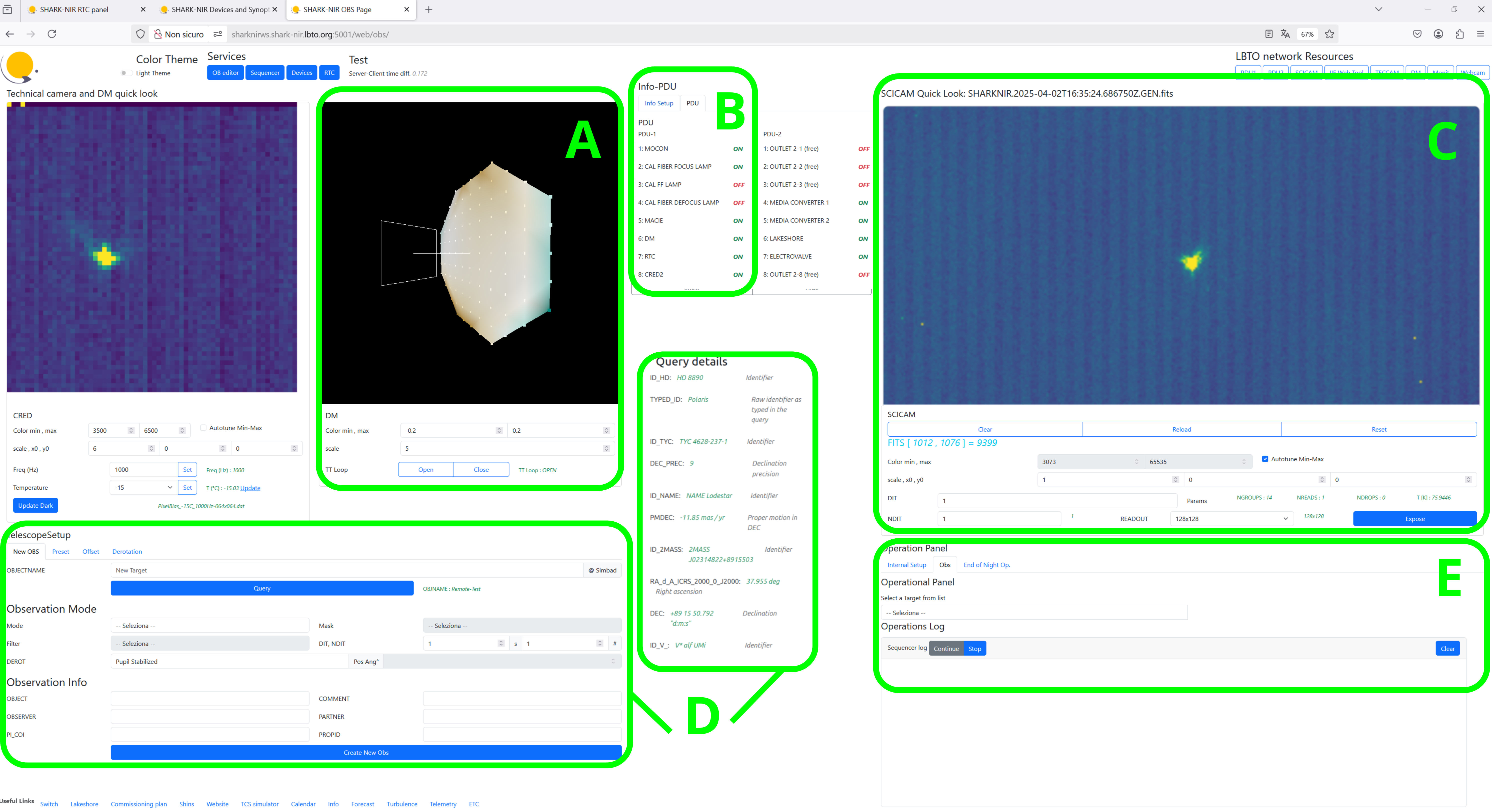}
  \caption{\label{fig:obs} Observation panel overview. Details in
    ``A'' (Fig.~\ref{fig:dm}),
    ``B'' (Fig.~\ref{fig:pdu}),
    ``C'' (Fig.~\ref{fig:scicam}),
    ``D'' (Fig.~\ref{fig:preset}),
    ``E'' (Fig.~\ref{fig:operations}).
  }
\end{figure}

OBs can be in principle prepared with any text editor, while the
sequencer can be executed as a standalone script in the instrument
workstation to launch them.

In order to ease the OBs preparation and execution, and in general to
easily use SHARK-NIR, we first developed a set of Representational
State Transfer Application Programming Interface \hl{(REST API)}
\cite{10.5555/932295} using the common HTTP verbs (\texttt{GET} to
retrieve a resource, \texttt{PUT} to modify it, \texttt{POST} to
create a new resource or start an operation, \texttt{DELETE} to delete
a resource or stop an operation.)  The \texttt{python} package
\texttt{Flask RestX} was used to build the API, in order to provide
automatic documentation generated by the \texttt{Swagger}
module\footnote{\url{https://swagger.io/}}, a set of tools that
simplify API development and documentation for users, teams, and
enterprises.

With these APIs
it is possible to create,
edit, delete, and launch OBs; check the value of the fixed parameters
of a given Template script; then monitor the execution of the OBs, and
directly interact with every single motor device, know its state and
position, interact with the RTC, and control the scientific camera.

\section{Web-based Graphical User Interfaces}
\label{sec:web-based-graphical}

On top of the REST APIs we developed a \texttt{Flask} web server and
several Graphical User Interfaces (GUIs) panels using the
\texttt{Bootstrap 5} framework.

\hl{Each panel is a web page. It is possible to open panels in
  separate browser tabs, or arrange them in separate browser windows,
  or a combination of both; as well as resize them with a reasonable
  responsiveness.}

Web pages use the REST API to trigger commands, and
the \texttt{Socket.IO} websocket library for the constant
communication of the SHINS status: when the web app is operative, a
client connection to \texttt{obs\_ctrl} is created and periodically
request for status information regarding the whole instrument.

A server class that stores status information in its attributes. A
single process asks the status of the instrument to \texttt{obs\_ctrl}
and updates the class attributes.  Web clients periodically receive
the values of these attributes using the websocket connection.

This ``REST for control + websocket for status'' approach has been
successfully tested\cite{2024SPIE13098E..0TR, 2022SPIE12186E..0PR} in
a $80\cm$ size telescope, and best practices have been applied
to SHARK-NIR.

\hl{In the following sections we present: the Devices and Synoptic
  panel of the instrument} (Fig.~\ref{fig:dev}), providing quick-look
  and control of individual motors and lamps; and the RTC engineering
  panel (Fig.~\ref{fig:rtc}), \hl{allowing quick-look and deep control
  of the TECCAM, the DM, and the RTC low-level functions. Then, after
  describing a separate, unreleased ``wizard'' tool for the
  preparation of the OBs, we present the general-purpose, ``user''
  Observation panel} (Fig.~\ref{fig:obs}), \hl{that was developed
  following the suggestions of the scientific team in order to provide
  the SHARK-NIR operator with a minimum but high-level set of
  functions and information. This panel also embeds a local version of
  the wizard, so that OBs can also be easily generated on-the-fly while
  operating at the telescope.

  A typical flow of a scientific observation of SHARK-NIR is then the
  following: the observations are prepared using the Wizard OB editor,
  then the folder with the OBs is uploaded on the instrument
  workstation. At the telescope, the user is expected opearating on
  the Observation panel to load and execute the OBs, keeping an eye on
  the Synoptic panel if necessary, and opening the RTC panel only for
  engineering purposes or debug.}

\subsection{Devices and Synoptic panel}
\label{sec:devic-synopt-panel}

\begin{figure}[t]
  \centering
  \begin{minipage}[t]{0.46\textwidth}
    \vspace{0pt}
    \centering
    \includegraphics[width=\textwidth]{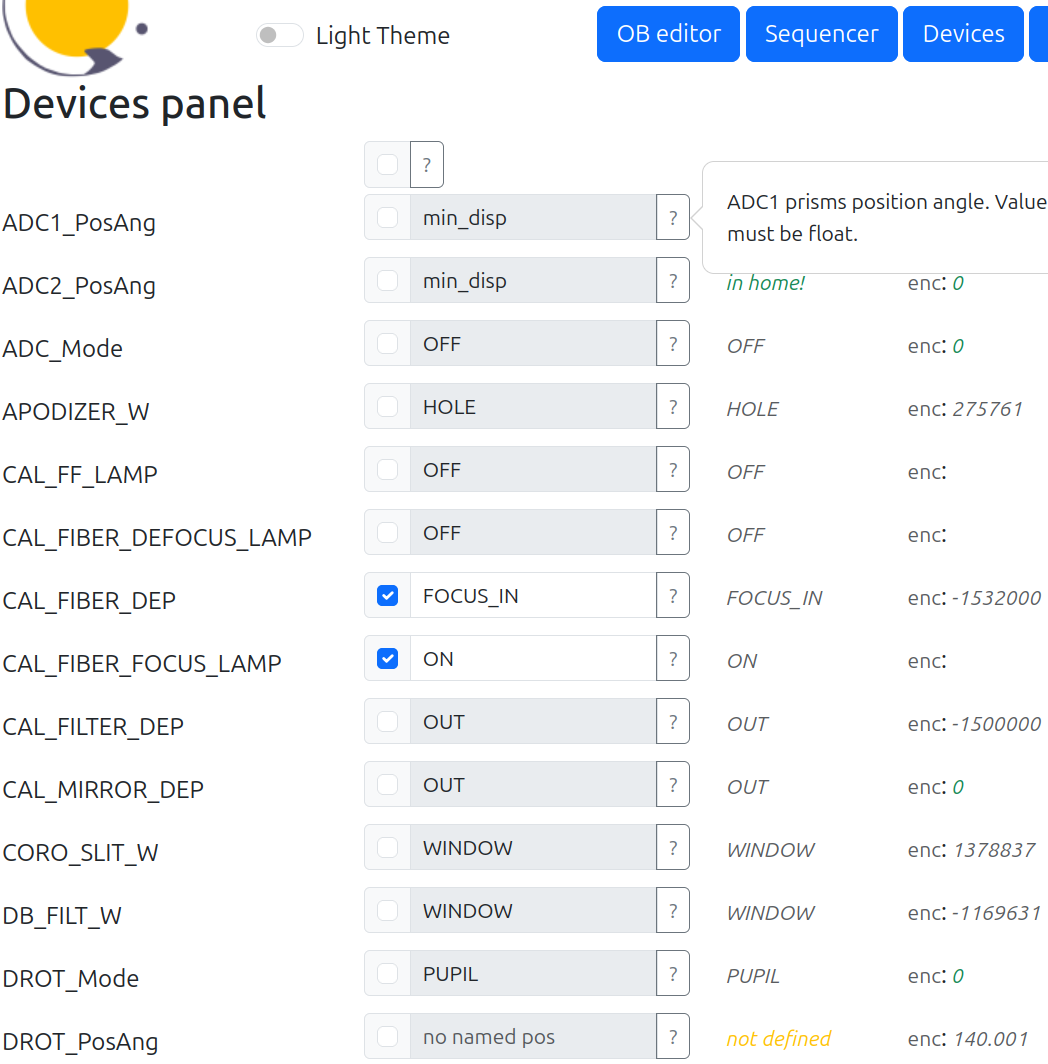}
    \caption{\label{fig:devices} Zoom on the Devices panel
      (see Fig.~\ref{fig:dev}). \hl{Two devices are selected in order to
      be moved together. The first ADC shows the help tooltip.} }
  \end{minipage}
  \hfill
  \begin{minipage}[t]{0.51\textwidth}
    \vspace{0pt}
    \centering
    \includegraphics[width=\textwidth]{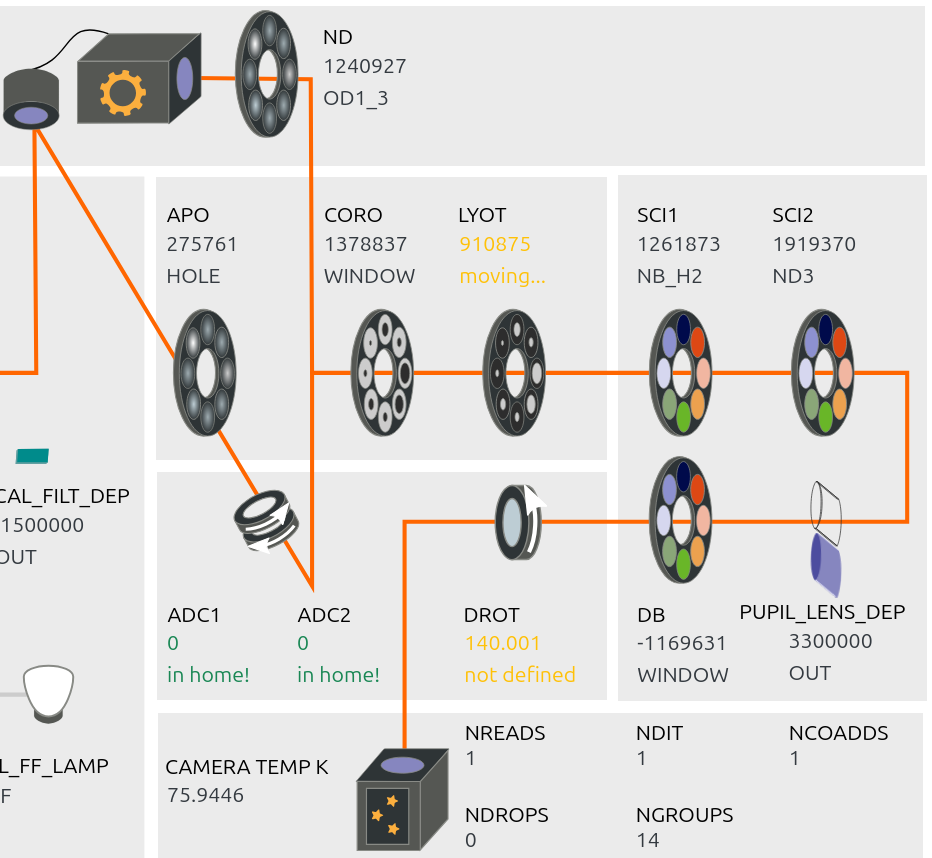}
    \caption{\label{fig:synoptic} Zoom on the Synoptic view (see
      Fig.~\ref{fig:dev}). \hl{The orange line shows the optical path
        receiving light}. }
  \end{minipage}
\end{figure}

Although the TaN framework provides a low-level, engineering GUI
developed in \texttt{Qt}, for each one of the \texttt{motion\_ctrl},
\texttt{calunit\_ctrl}, and \texttt{track\_ctrl} services, we decided
instead to develop a ``global'' Instrument Control panel for
SHARK-NIR, which we call Device and Synoptic panel (see
Fig.~\ref{fig:dev}).  We chose to go on with the APIs approach,
exposing HTTP methods to get and manage the status of devices. This
means that, even if this is mainly an engineering tool, these controls
pass through \texttt{obs\_ctrl} instead of operating at subsystem
level.

The left part of the panel (see a zoomed view in
Fig.~\ref{fig:devices}), labeled as ``Devices panel'', is used to
control and monitor the status of individual motors. Bulk operations
on multiple motors can be performed \hl{by selecting the corresponding
  checkboxes before clicking on the ``Move'' button. An additional
  checkbox on top of the list allows selecting/deselecting all devices
  with one click. Named position and position in unit of encoder steps
  are also shown. They are shown in green if in home position, in red
  if in error (for example when in contact with the limit switches),
  in yellow if moving or undefined, otherwhise in gray. }

The right part of the panel (see a zoomed view in
Fig.~\ref{fig:synoptic}), labeled as ``Synoptic panel'' shows a
graphical representation of the instrument's hardware devices, similar
to that shown in Fig.~\ref{fig:shark-opto}.  This helps the operator
to have a quick view of the setup. The Synoptic panel does not
reflects the real spatial displacement of the components; instead it
shows the devices as they are crossed by the optical path.
\hl{The optical path changes its color based on the setup, to enhance
whether the light source is coming from a specific calibration lamp or
from a source on sky: gray if not traveled by the light, otherwise
light orange}, and makes this path easier to understand. Rreadout
parameters of the SCICAM are also shown.

Moreover, the Synoptic Panel provides a base level of animation for
some devices; for instance: the shutter image can be opened or closed,
and the deployers' position changes based on the setup, and the lamps
can be on or off. Text fields show the motors position (expressed in
units of encoder steps), and their named position, if applicable,
following the same color schema of the Device section.

\subsection{RTC Panel}
\label{sec:rtc-panel}

\begin{figure}[t]
  \centering
  \begin{minipage}[t]{0.32\textwidth}
    \vspace{0pt}
    \centering
    \includegraphics[width=\textwidth]{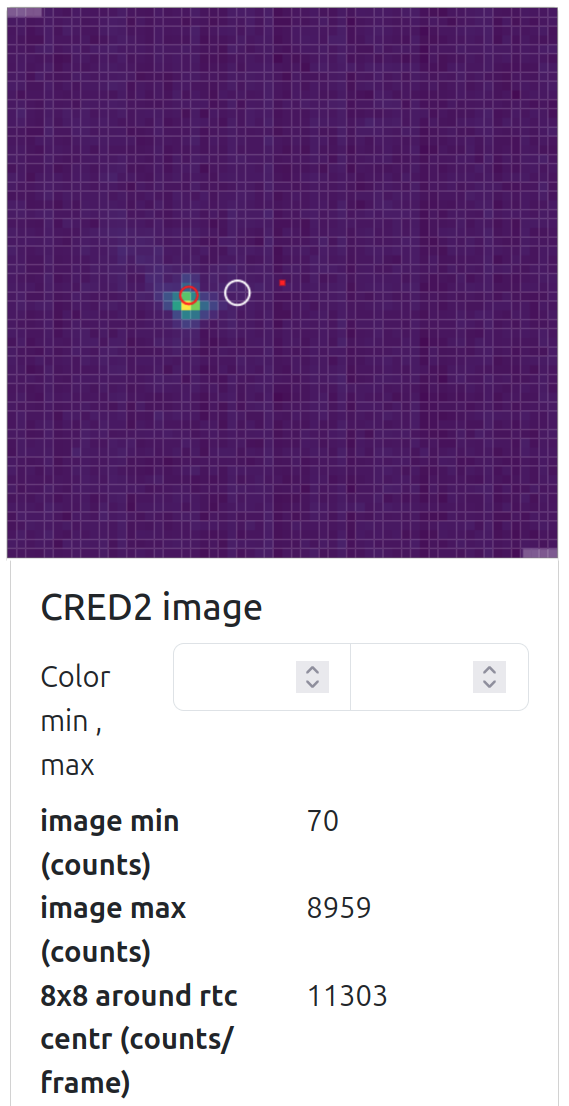}
    \caption{\label{fig:rtc-cred} Zoom on the CRED image (see
      Fig.~\ref{fig:rtc}). }
  \end{minipage}
  \hfill
  \begin{minipage}[t]{0.32\textwidth}
    \vspace{0pt}
    \centering
    \includegraphics[width=\textwidth]{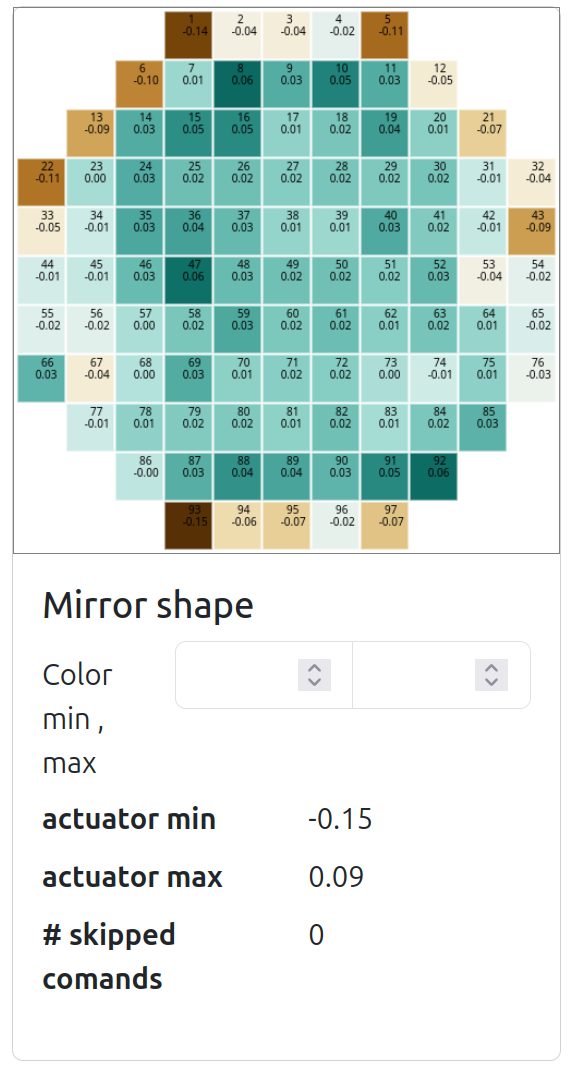}
    \caption{\label{fig:rtc-dm} Zoom on the mirror shape (see
      Fig.~\ref{fig:rtc}). }
  \end{minipage}
  \hfill
  \begin{minipage}[t]{0.32\textwidth}
    \vspace{0pt}
    \centering
    \includegraphics[width=\textwidth]{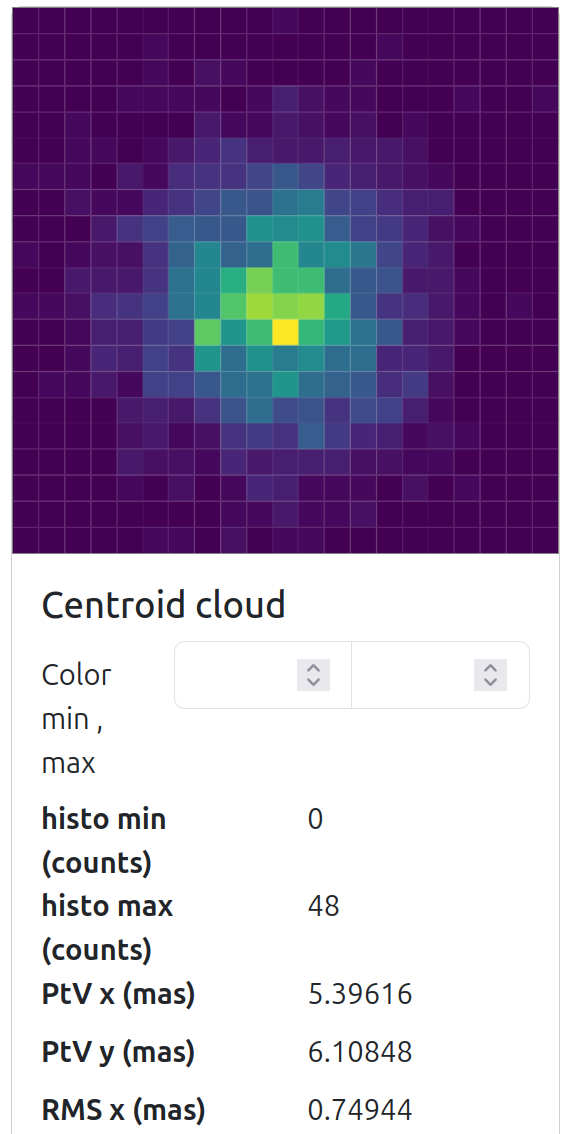}
    \caption{\label{fig:rtc-cloud} Zoom on centroid cloud (see
      Fig.~\ref{fig:rtc}). }
  \end{minipage}
\end{figure}

The RTC Panel (see Fig.~\ref{fig:rtc}) allows monitoring and fine
control of the wavefront control system of the instrument, and it was
specifically designed to support the operations during
Commissioning. \hl{The RTC panel was initially designed for quick-look
  and simple operations only, but as the assembly, integration,
  commissioning, and early science activities proceeded, more and more
  additions became necessary for testing, shaping it to a
  comprehensive engineering panel for the fine tuning and the debug of
  the SHARK-NIR Adaptive Optics subsystem}.
The panel is divided into two main areas: the Plot Area on top, and
the and the Control Area on bottom.

The Plot Area is based on a websocket communication established
directly between the upper \texttt{python}-based level and the
RTC. The communication is separately managed by the \texttt{socketio}
\texttt{python} (server-side) and \texttt{js} (client-side) package,
which implements a robust websocket system, periodically updating an
internal attribute with the diagnostic information received by the RTC
(reconstruction of the image on the TECCAM, measured centroid, DM
shape) by a separate \texttt{python} thread, called ``Python RTC
listener'' in Fig.~\ref{fig:shark-stack}, \hl{which was adapted from
  an existing python script provided by Microgate}. This allows us to
reduce the request load on \texttt{obs\_ctrl}.

Fig~\ref{fig:rtc} shows four plots, each one representing a
fundamental diagnostic information, namely:
\begin{description}
\item[C-RED2 image:] a TECCAM Plot (see Fig.~\ref{fig:rtc-cred}), a
  representation of the active region of the technical camera, to let
  the instrument operator always check the position of the PSF and to
  monitor the tracking status of the loop. In this plot is also shown
  as a white circle the point corresponding to the center of the
  coronagraphic mask.  Moreover, a red circle shows the point that the
  RTC calculates as the PSF centroid. Finally, a red dot shows the
  point where the RTC moves the PSF while operating in closed loop. In
  Fig.~\ref{fig:rtc-cred}, \hl{the loop is open, and the PSF is still
    not placed behind the mask};
\item[Mirror shape:] a graphical visualization \hl{of the stroke value} of
  the 97 actuators of the DM (see Fig.~\ref{fig:rtc-dm}), drawn as a
  set of squares in their corresponding position behind the
  mirror. The level of stroke of each actuator is represented using a
  color scale that ranges from the min to max value of the current
  configuration;
\item[Mirror shape without flat:] a second reproduction of the
  actuator, similar to the previous one, but without taking into
  account the internal-\emph{static} NCPA (see
  Sect.~\ref{sec:ncpa-rotation}), that are called ``Flat'' in the RTC
  panel. This third plot is mainly used for the definition and
  optimization of the external-\emph{rotating} NCPA, that are called
  simply ``NCPA'' in the panel;
\item[Centroid cloud:] a statistical representation of the centroid
  position in the last $N$ TECCAM images (see
  Fig.~\ref{fig:rtc-cloud}), helping to monitor internal loop
  performances and residual jitter.
\end{description}

Below each plot, a custom color range can be selected, and the values
are updated on the plots at every refresh of the information.  Basic
statistical information are also shown and continuously updated.

\begin{figure}[t]
  \centering
    \includegraphics[width=0.91\textwidth]{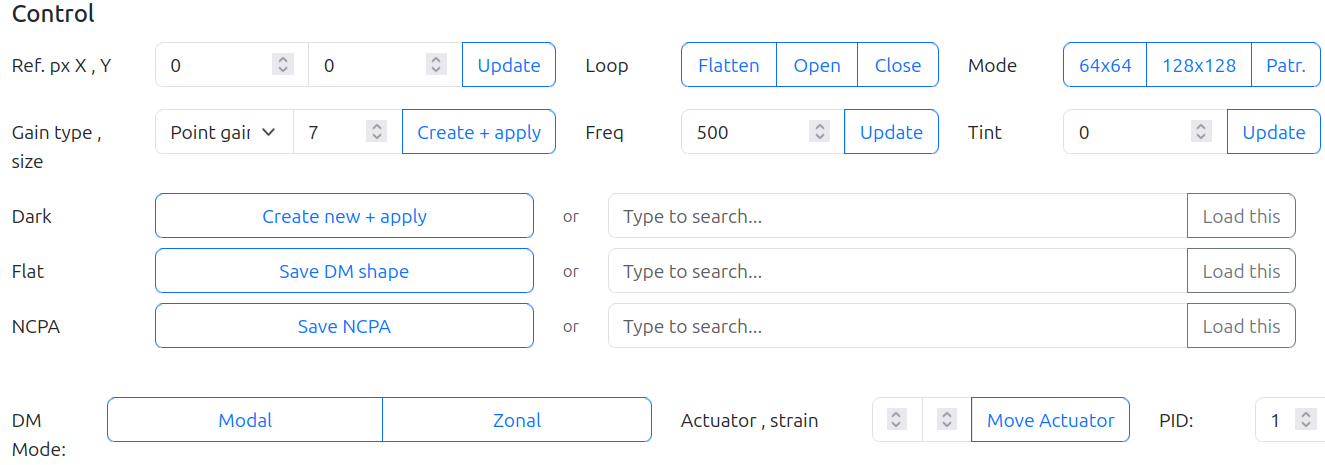}
    \hfill
    \includegraphics[width=0.08\textwidth]{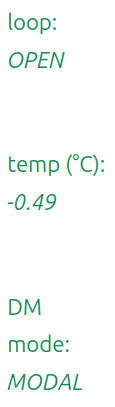}
    \caption{\label{fig:rtc-control} Zoom on the Control area (left)
      and on the status information (right, green) of the RTC panel
      (see Fig.~\ref{fig:rtc}). }
\end{figure}

The Control Area, on the other hand, allows to send commands to
\texttt{tiptilt\_ctrl} through \texttt{obs\_ctrl} via REST API to the
RTC.  Fig.~\ref{fig:rtc-control} shows a selection of the the main
control operations, including: setting TECCAM parameters (as
frequency and, integration time), define the gain
mask (a sub-area on the TECCAM to optimize the computation of the PSF
centroid for close loop operations), load the dark background of the
TECCAM, and the NCPA.

The RTC calculates the PSF centroid on a $ 64\time 64 \pixel$
subaperture of the TECCAM by default. This region is used while
operating the DM in closed loop.  Moreover, for test purposes, the
TECCAM can also operate full-frame as ``patrol camera'', \hl{without
  the possibility to operate for tip-tilt correction}.  TECCAM Mode
buttons allow to switch between tip-tilt correction mode, and patrol
camera mode.

Other commands allow to open/close the internal fast tip-tilt loop,
set the reference position for the PSF in closed loop, command single
relative tip-tilt commands to the DM.  Moreover, the control panel
also allows to change the internal interaction matrix of the RTC,
being able to switch between modal mode to zonal mode, with the latter
that allows to singly command the actuators; and to change the PID of
the DM.

The right side of the control area (see an example on right of
Fig.~\ref{fig:rtc-control}, in green) collects all the status
information, collected by websocket from \texttt{obs\_ctrl} and
updated every $0.5\second$.

\subsection{Wizard editor}
\label{sec:wizard}

\begin{figure}[t]
  \centering
  \begin{minipage}[t]{0.60\textwidth}
    \vspace{0pt}
    \centering
  \includegraphics[width=\textwidth]{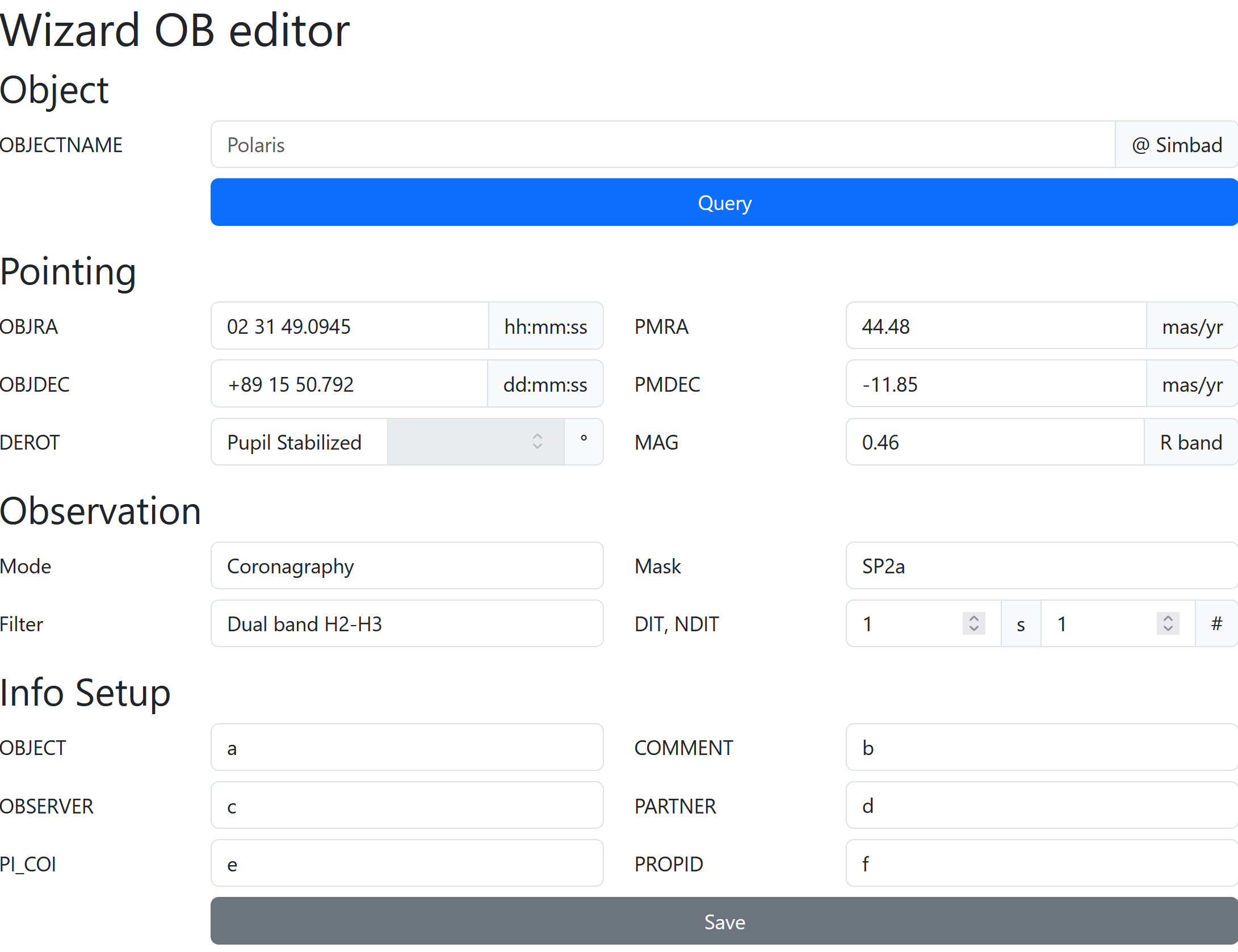}
  \caption{\label{fig:wizard} \hl{Wizard OB editor, currently for
      internal use of the SHARK-NIR team. It produces a folder with a
      complete set of OBs in XML format, including acquisition
      procedure, calibrations relative to the chosen configuration,
      and observation sequence. }}
  \end{minipage}
  \hfill
  \begin{minipage}[t]{0.38\textwidth}
    \vspace{0pt}
    \centering
    \includegraphics[width=\textwidth]{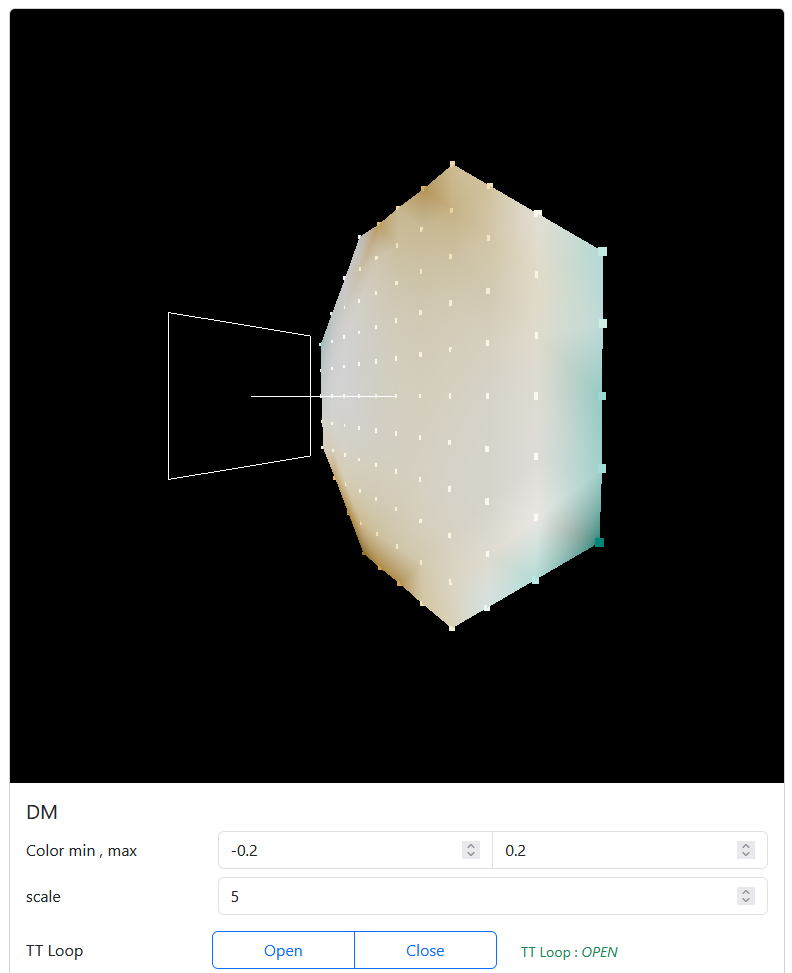}
    \caption{\label{fig:dm} Zoom on the 3D DM quick look section in the
      Observation panel (see Fig.~\ref{fig:obs}), with commands to
      open and close the tip-tilt loop .}
  \end{minipage}
\end{figure}

\hl{ At the end of the Commissioning phase, we brainstormed all the
technologic and scientific team, to condensate the typical use of
SHARK-NIR into an ``operation flow''.
This lead to the creation of two tools: the first, presented in this
section, is an external ``wizard'' editor} (see Fig.~\ref{fig:wizard}),
\hl{allowing astronomers to preapare in advance, and in separate server
with respect to the instrument workstation (i.e. currently in an
external server accessible by the SHARK-NIR team only), a complete set
of OBs based on the Instrument Mode and a minimal set of information,
including of course the scientific target.}

First, the target coordinates can be retrieved by a Preset query that
searches the object on the
Simbad\footnote{\url{http://simbad.u-strasbg.fr/simbad/}}
catalog. \hl{A further software development will also handle
  non-sidereal targets, which are currently handled directly at LBT by
  the telescope operator.}

The  query automatically populates the target coordinates.
The user completes this section by adding the additional LBT pointing
flags, including the use of secondary mirror's Adaptive Optics, the
Telescope mode (i.e. which LBT side is leading the observation), and
deotator mode, as well as the flags controlling the Binocular mode.

Then, the user fills the rest of the form, choosing the SHARK-NIR
instrument mode (Coronagraphy, Direct Imaging, or Long Slit
Spectroscopy).  Depending on the chosen mode, then a set of
coronagraphic masks, no masks, or a set of spectroscopy masks are
available. Depending on both the Observation Mode and the mask,
available combination of filters or no filters are available.
The last information is the derotation type, which can be
pupil-stabilized (no instrument derotation), a field-stabilized mode
that maximizes the derotation angle, and a field-stabilized mode at a
custom position angle.
Finally, typical Observation information (such as proposal ID, name of
the Principal Investigator) is filled by the astronomer.
\hl{By clicking on the button ``Save'', a complete set of
Observation Blocks based on the previous information are created and
saved on a new directory.  These OBs include: the target acquisition
sequence, the alignment procedures corresponding to the chosen
combination of Instrument Mode and Mask, the calibration OBs, and of
course the scientific observation sequence.

Currently, the wizard is for internal use of the SHARK-NIR team, and
the directories with the OBs are uploaded on \texttt{sharknirws}
manually. It is foreseen to make this tool publicly available.}

\subsection{Observation panel}
\label{sec:observation-panel}

\begin{figure}[t]
  \centering
  \begin{minipage}[t]{0.29\textwidth}
    \vspace{0pt}
    \centering
    \includegraphics[width=\textwidth]{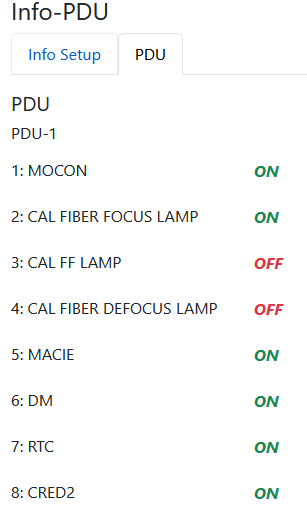}
    \caption{\label{fig:pdu} Zoom of the central tabs in the
      Observation panel (see Fig.~\ref{fig:obs}). Info setup (in
      background), allowing to edit the complimentary OB information;
      and read-only PDU informations (in foreground). }
  \end{minipage}
  \hfill
  \begin{minipage}[t]{0.70\textwidth}
    \vspace{0pt}
    \centering
    \includegraphics[width=\textwidth]{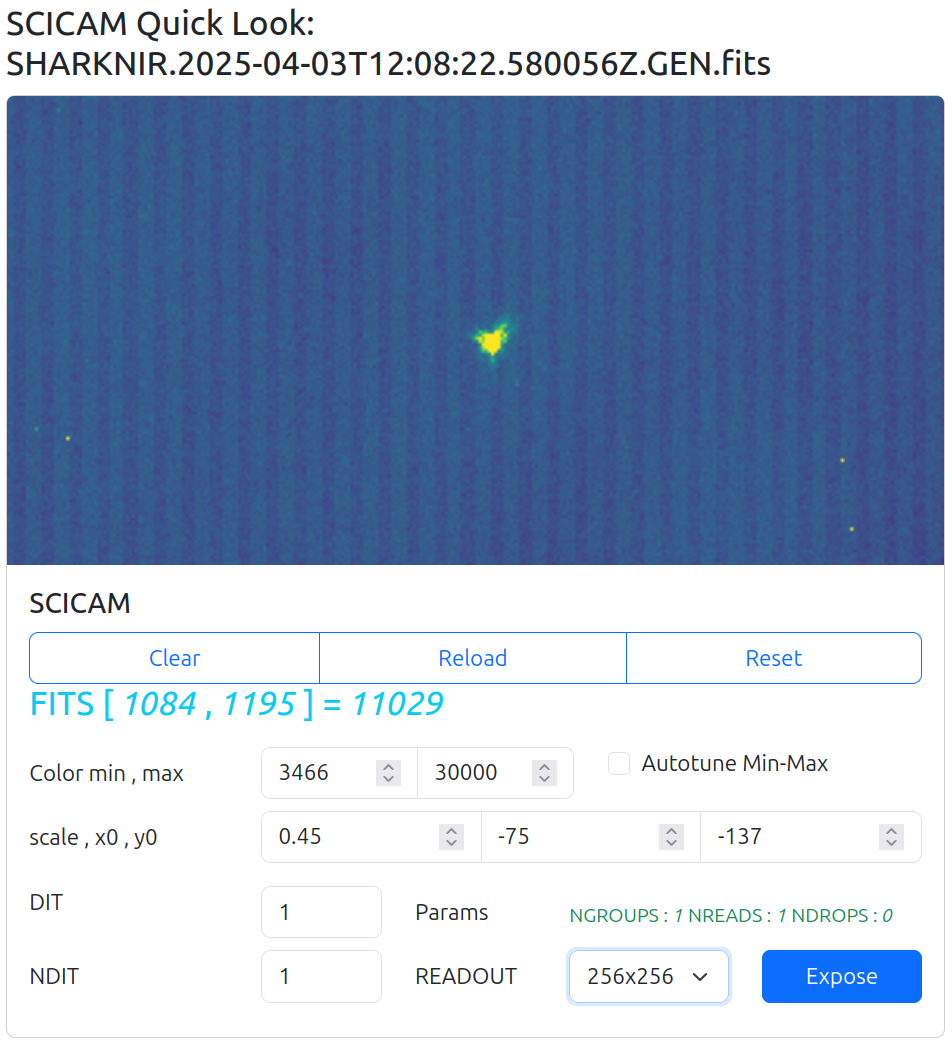}
    \caption{\label{fig:scicam} Zoom on the SCICAM control section of
      the Observation panel (see Fig.~\ref{fig:obs}).}
  \end{minipage}
\end{figure}

\hl{The second tool developed after the brainstorming, presented in
  this section, is the SHARK-NIR ``Observation panel'' , containing
  the minimal information to control and monitor the instrument at the
  telescope, and to operate basic troubleshooting.
  This step reduced the degrees of freedom given by the engineering
  panels and by the parameters of the Templates scripts. This allowed
  developing a high-level interface mainly for scientific use.
  The panel also embeds a slightly modified version of the ``wizard''
  editor, in order to create on-the-fly an observation directly from
  the panel, if necessary, while operating at LBT.}

\begin{figure}[t]
  \centering
  \includegraphics[width=\textwidth]{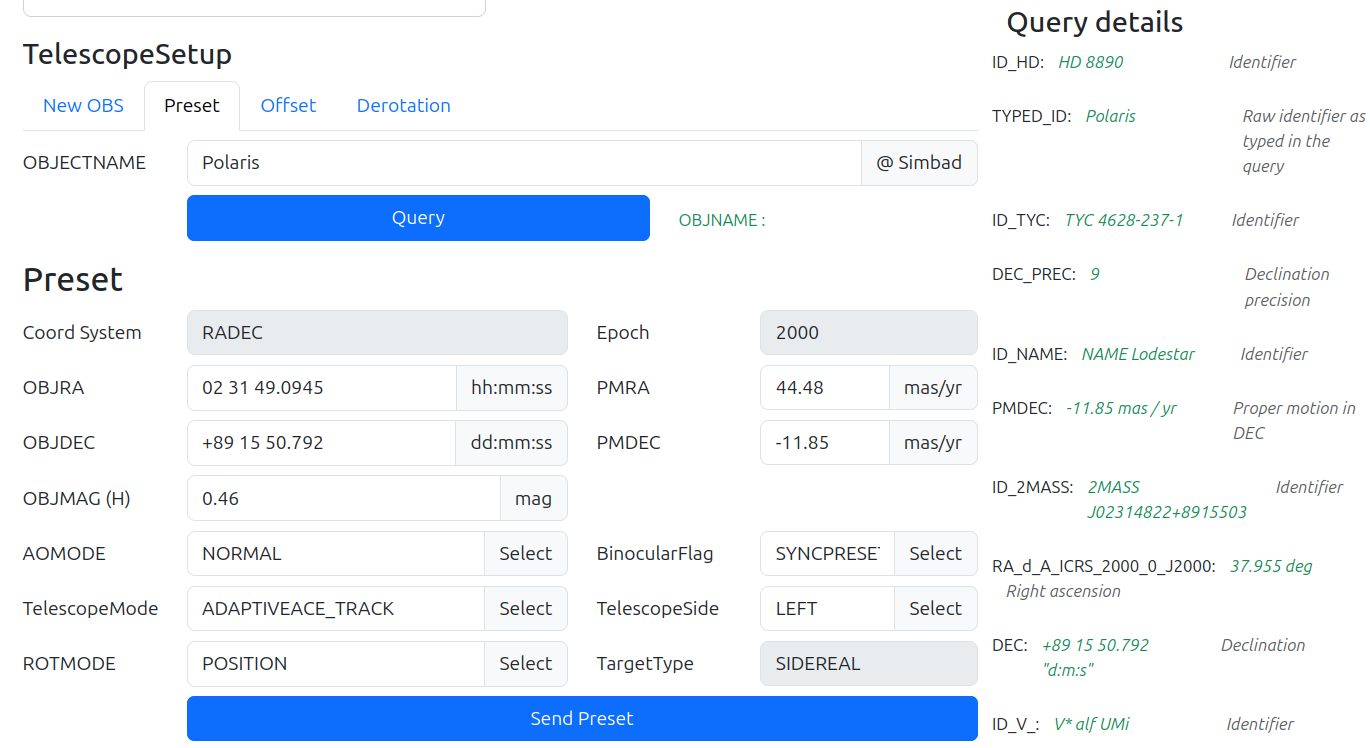}
  \caption{\label{fig:preset} \hl{Zoom on the lower tabs section of
      the Observation panel} (see Fig.~\ref{fig:obs}).  On the right,
    the details of the queried object, retrieved from Simbad. }
\end{figure}

We describe in detail this panel as an effort to make this paper
useful also as a short version of the user manual of the instrument.

The top part of the panel (Fig.~\ref{fig:obs}) shows:
\begin{description}
\item[Technical Camera quick look:] a WebGL image, and its respective
  color range and pan-zoom controls, in a way similar to what provided
  by the RTC panel.  It allows tuning the temperature, the frequency,
  and updating the dark, which is subtracted to improve PSF
  centroid Signal-to-Noise ratio.
\item[DM quick look:] a WebGL 3D image, and its respective color range
  (see Fig.~\ref{fig:dm}).  In this case, scale controls are used to
  enhance the DM stroke. The user can open and close the
  tip-tilt loop as quick debug feature.
\item[Info-PDU:] it provides two tabs (see Fig.~\ref{fig:pdu}). The
  first shows the complimentary setup information; the second allows
  bare monitoring of the state of the PDU outlets that control the
  \emph{Lamps} devices described in
  Sect.~\ref{sec:control-electronics}, as well as other outlets
  powering the instrument.
\item[SCICAM quick look:] it shows the latest image taken with the
  Scientific Camera, and its respective color range and pan-zoom
  controls (see Fig.~\ref{fig:scicam}). It also allows exposing test
  snapshot images by specifying the Detector Integration Time, the
  number of sub-integrations, and the readout region from a
  pre-defined list of Region of Interests. Pan and zoom on the images
  are also available by mouse drag and wheel events.  As in the other
  panels, the status information in green are retrieved via websocket
  from \texttt{obs\_ctrl}.
\end{description}

\hl{As in the RTC Engineering panel, this top part of the Observation
  panel is dedicated to monitor and quick tests. The bottom part of
  the panel shows:}

\begin{description}
\item[Local wizard and telescope control tabs:] Fig.~\ref{fig:preset}
  \hl{background ``New OBS'' tab hides a local replica of the wizard,
    that is used to create on-the-fly a set of OBs, if necessary. It
    also populates the foreground ``Preset'' tab, that allows
    modifying the coordinates and the telescope observation mode. This
    tab is also used independently to communicate with the telescope,
    if necessary. The Send Preset button allows to actually point the
    telescope. Other tabs allow fine-tuning the telescope offset and
    or change the SHARK-NIR derotation mode.  }
\item[Operation section:] \hl{is the top level control of SHARK-NIR.
    The user retrieves from a dropdown menu the OBs directories
    previously generated by the wizard OB editor and uploaded on the
    instrument workstation, or generated on-the-fly from the local
    wizard} (\texttt{2025-03-18\_Polaris\_DI} in
  Fig.~\ref{fig:operations}).
  \hl{ The OBs are loaded in a numbered, button-based sequencer, so
    that the operator can run them one by one, or repeat a given step
    changing common parameters if necessary.  A tooltip shows the
    template help.}  Fig.~\ref{fig:operations} also shows the
  ``Internal Setup'' and ``End of Night Op.'' tabs, containing OBs
  with leading and trailing common operations.

  Operations logs are shown in a box that reads the instrument log
  files and updates it each $0.5 \second$.

\end{description}

If there is a need to access less common parameters for peculiar tests
or conditions, then a separate panel called ``OB editor'', that it is
not described here, but is available by clicking the ``edit'' button
of each operation step.  Developed for the AIT phase, is thought of as
a simplified version of ESO P2 system and also shows the reference
setup in disabled boxes.

\section{Scientific results}

\begin{figure}[t]
  \centering
  \begin{minipage}[t]{0.60\textwidth}
    \vspace{0pt}
    \centering
    \includegraphics[width=\textwidth]{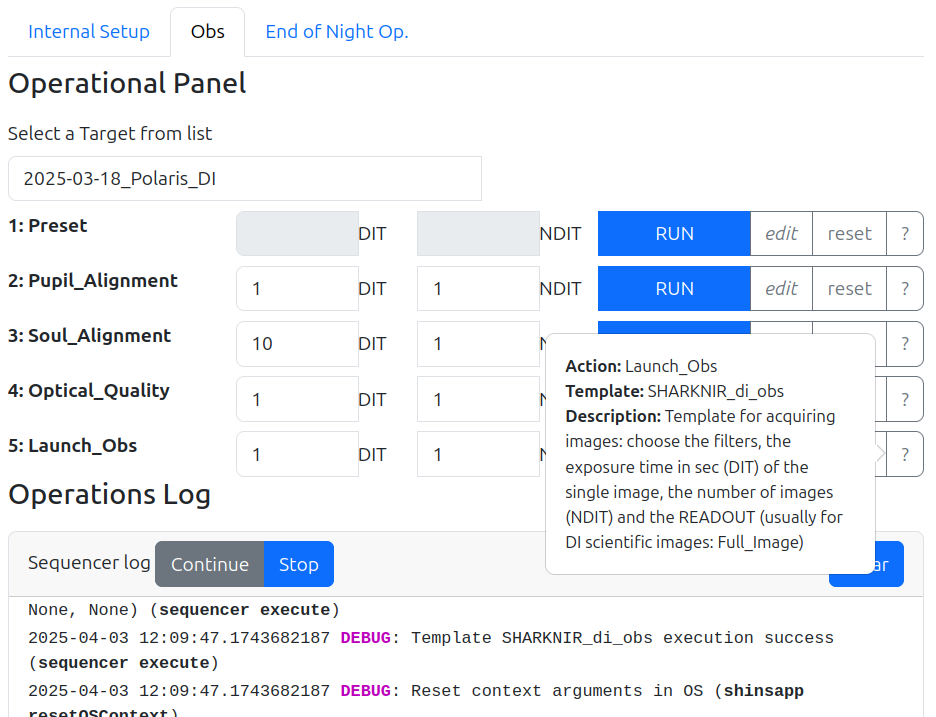}
    \caption{\label{fig:operations} \hl{Zoom on the operational tabs
        of the Observation panel} (see Fig.~\ref{fig:obs}), and to the
      Operations Log. See Sect.~\ref{sec:observation-panel} for
      details.}
  \end{minipage}
  \hfill
  \begin{minipage}[t]{0.39\textwidth}
    \vspace{0pt}
    \centering
  \includegraphics[width=\textwidth]{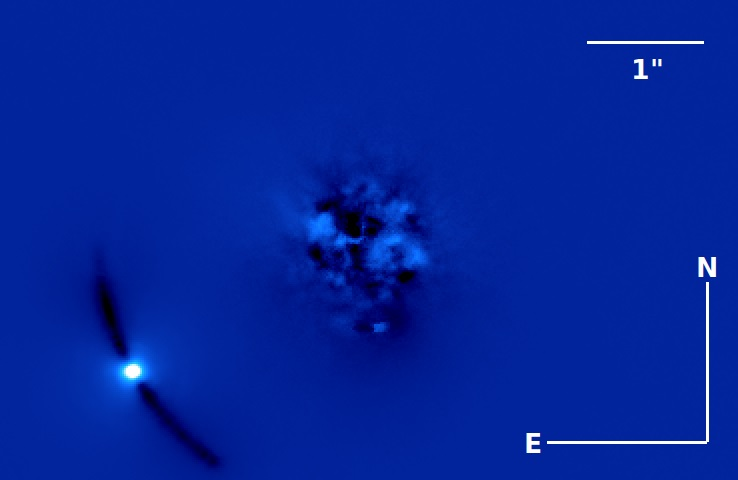}
  \caption{\label{fig:hip-36277} \hl{Final image of HIP36277 observed
      on 2024-02-21 with SHARK-NIR using the Gaussian coronagraph and
      the broad H-band filter (BB\_H). The total rotation of the field
      of view was $33.6\degree$ while the total exposure time on the
      target was $5660.5\second$ with single DITs of $61.5\second$. A
      bright stellar companion is visible South-East from the star
      while a fainter candidate companion is visible at shorter
      separation ($0.625\arcsecond$) from the star.}}
  \end{minipage}
\end{figure}

In its Early Science phase, SHARK-NIR obtained its first scientific
high contrast image results in the $H$ band, using the Gaussian
coronagraphic mask, which is one of the installed coronagraph together
with Shaped Pupil masks and a Four Quadrant Phase
Mask\cite{2000PASP..112.1479R}. Observations, obtained in
pupil-tabilized mode, have been combined with observations with
LBTI/LMIRCam in $L^\prime$, all involved instruments operating in
parallel at LBT.

The first result~\cite{2025A&A...693A..81B} is the the contribution to
the characterization of an eccentric giant planet detected around the
star \textsf{HD~57625}. LBT observations have been combined with
available SOPHIE radial velocities and Hipparcos-Gaia Proper Motion
Anomaly (PMa) measurements, finding the following parameters: a
$8.43^{+1.1}_{-0.91}M_{Jup}$ planetary companion mass, an orbital
separation of $5.70^{+0.14}_{-0.13}\rm au$ and an eccentricity of
$0.52^{+0.04}_{-0.03}$.

The second result~\cite{2025MNRAS.536.1455M} is a deep imaging study
of three accelerating stars: \textsf{HIP~11696}, for which the
analysis suggests the presence of a companion with a mass between $4$
and $16M_{Jup}$. at a $2.5$--$28\rm au$ separation;
\textsf{HIP~47110}, for which the study finds a $2$--$10M_{Jup}$
companion at $3$ to $30\rm au$; and finally \textsf{HIP~36277},
suggesting a second candidate companions in addition to the first,
detected by SHARK-NIR observations (see Fig.~\ref{fig:hip-36277}) and
already confirmed by Gaia.

\section{Conclusion}
\label{sec:conclusion}

We presented SHINS, the Instrument Control Software of SHARK-NIR, a
new instrument at LBT for high contrast imaging.

SHINS is principally based on the TaN framework to control most of its
subsystems, while exploits SNMP and INDI protocols to control the
scientific camera and the calibration lamps, respectively.  Its
central component, the Observation Control Software, is developed on
the ICE framework, all being based on \texttt{C++}. The central
components also implement safety procedures and NCPA correction
management, which is one of the most critical tasks of the wavefront
control system of the instrument.

High-level components are developed in \texttt{python}, and include a
sequencer for Template scripts called by XML OBs. Even if the
sequencer can be called by terminal, SHINS users can operate thanks to
a set of web-based GUIs for observation and engineering purposes, that
make use of REST API for control and websockets for monitoring. WebGL
is used for quick-look of scientific and technical images, as well as
to monitor the shape of the internal DM.

Users can easily plan their operations with a ``wizard'' tool that
help the configuration of OBs for all steps of the observation.

Early science tests demonstrated that SHINS allows to operate
SHARK-NIR exploiting its full potential, allows significant results in
the exoplanetary field, and it is ready for regular operations.

\section*{Disclosures}

The authors declare there are no financial interests, commercial
affiliations, or other potential conflicts of interest that have
influenced the objectivity of this research or the writing of this
paper.

\section*{Code and Data}

Data sharing is not applicable to this article.

\acknowledgments

We thank Tom Herbst from MPIA-Heidelberg, and the LINC-NIRVANA team,
for sharing part of their instrument control SW to operate the
motorized axis of SHARK-NIR.
We also express our appreciation to NASA and Marcia Rieke, the
Principal Investigator of JWST/NIRCam, for granting us the opportunity
to utilize one of the NIRCam spare detectors as the primary detector
for the SHARK-NIR scientific camera.
Observations have benefited from the use of ALTA Center
(\url{alta.arcetri.inaf.it}) forecasts performed with the
Astro-Meso-Nh model. Initialization data of the ALTA automatic
forecast system comes from the General Circulation Model (HRES) of the
European Centre for Medium Range Weather Forecasts.
DR acknowledges fundings from ``MINI-GRANTS (2023) di RSN5'',
\texttt{C93C23008400001}.
The LBT is an international collaboration among institutions in the
United States, Italy, and Germany. The LBT Corporation partners are:
The University of Arizona on behalf of the Arizona University system;
Istituto Nazionale di Astrofisica, Italy; LBT
Beteiligungsgesellschaft, Germany, representing the Max Planck
Society, the Astrophysical Institute Potsdam, and Heidelberg
University; The Ohio State University; The Research Corporation, on
behalf of The University of Notre Dame, University of Minnesota and
University of Virginia.

\section*{Biography}

The first author is Technologist at the INAF - Osservatorio
astronomico di Padova. He is involved in Instrument Control Software
of ESO and LBT projects, as well as in the development of small
observatories for science and outhreach purposes. His interests span
from web user interfaces for telescopes, to extrasolar planets
observations.


\bibliography{biblio} 

\begin{thebibliography}{10}

\bibitem{2016SPIE.9909E..3VP}
E.~{Pinna}, S.~{Esposito}, P.~{Hinz}, {\em et~al.}, ``{SOUL: the Single
  conjugated adaptive Optics Upgrade for LBT},'' in {\em Adaptive Optics
  Systems V},  E.~{Marchetti}, L.~M. {Close}, and J.-P. {V{\'e}ran}, Eds., {\em
  Society of Photo-Optical Instrumentation Engineers (SPIE) Conference Series}
  {\bf 9909}, 99093V  (2016).

\bibitem{2016SPIE.9908E..32P}
F.~{Pedichini}, F.~{Ambrosino}, M.~{Centrone}, {\em et~al.}, ``{The V-SHARK
  high contrast imager at LBT},'' in {\em Ground-based and Airborne
  Instrumentation for Astronomy VI},  C.~J. {Evans}, L.~{Simard}, and
  H.~{Takami}, Eds., {\em Society of Photo-Optical Instrumentation Engineers
  (SPIE) Conference Series} {\bf 9908}, 990832  (2016).

\bibitem{2016SPIE.9911E..27V}
V.~{Viotto}, J.~{Farinato}, D.~{Greggio}, {\em et~al.}, ``{SHARK-NIR system
  design analysis overview},'' in {\em Modeling, Systems Engineering, and
  Project Management for Astronomy VI},  G.~Z. {Angeli} and P.~{Dierickx},
  Eds., {\em Society of Photo-Optical Instrumentation Engineers (SPIE)
  Conference Series} {\bf 9911}, 991127  (2016).

\bibitem{2016SPIE.9909E..31F}
J.~{Farinato}, F.~{Bacciotti}, C.~{Baffa}, {\em et~al.}, ``{SHARK-NIR: from
  K-band to a key instrument, a status update},'' in {\em Adaptive Optics
  Systems V},  E.~{Marchetti}, L.~M. {Close}, and J.-P. {V{\'e}ran}, Eds., {\em
  Society of Photo-Optical Instrumentation Engineers (SPIE) Conference Series}
  {\bf 9909}, 990931  (2016).

\bibitem{2015IJAsB..14..365F}
J.~{Farinato}, C.~{Baffa}, A.~{Baruffolo}, {\em et~al.}, ``{The NIR arm of
  SHARK: System for coronagraphy with High-order Adaptive optics from R to K
  bands},'' {\em International Journal of Astrobiology} {\bf 14}, 365--373
  (2015).

\bibitem{2018SPIE10701E..2BC}
E.~{Carolo}, D.~{Vassallo}, J.~{Farinato}, {\em et~al.}, ``{SHARK-NIR
  coronagraphic simulations: performance dependence on the Strehl ratio},'' in
  {\em Optical and Infrared Interferometry and Imaging VI},  M.~J.
  {Creech-Eakman}, P.~G. {Tuthill}, and A.~{M{\'e}rand}, Eds., {\em Society of
  Photo-Optical Instrumentation Engineers (SPIE) Conference Series} {\bf
  10701}, 107012B  (2018).

\bibitem{2024SPIE13096E..3TC}
P.~{Cerpelloni}, E.~{Carolo}, G.~{Umbriaco}, {\em et~al.}, ``{Unravelling the
  performance of the SHARK-NIR Four Quadrant Phase Mask in a controlled
  environment},'' in {\em Ground-based and Airborne Instrumentation for
  Astronomy X},  J.~J. {Bryant}, K.~{Motohara}, and J.~R.~D. {Vernet}, Eds.,
  {\em Society of Photo-Optical Instrumentation Engineers (SPIE) Conference
  Series} {\bf 13096}, 130963T  (2024).

\bibitem{2024SPIE13096E..1WB}
D.~{Barbato}, J.~{Farinato}, A.~{Baruffolo}, {\em et~al.}, ``{SHARK-NIR
  commissioning and early science runs},'' in {\em Ground-based and Airborne
  Instrumentation for Astronomy X},  J.~J. {Bryant}, K.~{Motohara}, and
  J.~R.~D. {Vernet}, Eds., {\em Society of Photo-Optical Instrumentation
  Engineers (SPIE) Conference Series} {\bf 13096}, 130961W  (2024).

\bibitem{2022SPIE12187E..09B}
M.~{Bergomi}, L.~{Marafatto}, E.~{Carolo}, {\em et~al.}, ``{SHARK-NIR: from
  design to installation, ready to dive into first light},'' in {\em Modeling,
  Systems Engineering, and Project Management for Astronomy X},  G.~Z. {Angeli}
  and P.~{Dierickx}, Eds., {\em Society of Photo-Optical Instrumentation
  Engineers (SPIE) Conference Series} {\bf 12187}, 1218709  (2022).

\bibitem{2022SPIE12185E..8IV}
D.~{Vassallo}, M.~{Bergomi}, E.~{Carolo}, {\em et~al.}, ``{Laboratory
  demonstration of focal plane wavefront sensing using phase diversity: a way
  to tackle the problem of NCPA in SHARK-NIR: Part II: new characterization
  tests and alternative wavefront sensing strategies},'' in {\em Adaptive
  Optics Systems VIII},  L.~{Schreiber}, D.~{Schmidt}, and E.~{Vernet}, Eds.,
  {\em Society of Photo-Optical Instrumentation Engineers (SPIE) Conference
  Series} {\bf 12185}, 121858I  (2022).

\bibitem{2022SPIE12185E..6WU}
G.~{Umbriaco}, D.~{Vassallo}, J.~{Farinato}, {\em et~al.}, ``{Deformable lens
  for testing the performance of focal plane wavefront sensing using phase
  diversity},'' in {\em Adaptive Optics Systems VIII},  L.~{Schreiber},
  D.~{Schmidt}, and E.~{Vernet}, Eds., {\em Society of Photo-Optical
  Instrumentation Engineers (SPIE) Conference Series} {\bf 12185}, 121856W
  (2022).

\bibitem{2022SPIE12185E..22F}
J.~{Farinato}, A.~{Baruffolo}, M.~{Bergomi}, {\em et~al.}, ``{SHARK-NIR, ready
  to ``swim'' in the LBT Northern Hemisphere ``ocean''},'' in {\em Adaptive
  Optics Systems VIII},  L.~{Schreiber}, D.~{Schmidt}, and E.~{Vernet}, Eds.,
  {\em Society of Photo-Optical Instrumentation Engineers (SPIE) Conference
  Series} {\bf 12185}, 1218522  (2022).

\bibitem{2022SPIE12184E..3VM}
L.~{Marafatto}, E.~{Carolo}, G.~{Umbriaco}, {\em et~al.}, ``{SHARK-NIR on its
  way to LBT},'' in {\em Ground-based and Airborne Instrumentation for
  Astronomy IX},  C.~J. {Evans}, J.~J. {Bryant}, and K.~{Motohara}, Eds., {\em
  Society of Photo-Optical Instrumentation Engineers (SPIE) Conference Series}
  {\bf 12184}, 121843V  (2022).

\bibitem{2025A&A...693A..81B}
D.~{Barbato}, D.~{Mesa}, V.~{D'Orazi}, {\em et~al.}, ``{A multi-technique
  detection of an eccentric giant planet around the accelerating star HD
  57625},'' {\em \aap} {\bf 693}, A81  (2025).

\bibitem{2025MNRAS.536.1455M}
D.~{Mesa}, R.~{Gratton}, V.~{D'Orazi}, {\em et~al.}, ``{Deep imaging of three
  accelerating stars using SHARK-NIR and LMIRCam at LBT},'' {\em \mnras} {\bf
  536}, 1455--1466  (2025).

\bibitem{lorenzetto2024}
A.~Lorenzetto, D.~Ricci, F.~Laudisio, {\em et~al.}, ``{SHARK-NIR instrument
  control software: new features and tools approaching science runs},'' in {\em
  Software and Cyberinfrastructure for Astronomy VIII},  J.~Ibsen and
  G.~Chiozzi, Eds.,  {\bf 13101}, 131013S, International Society for Optics and
  Photonics, SPIE  (2024).

\bibitem{2022SPIE12189E..20R}
D.~{Ricci}, F.~{Laudisio}, S.~{Chavan}, {\em et~al.}, ``{Improvements to SHINS,
  the SHARK-NIR instrument software, during the AIT phase},'' in {\em Software
  and Cyberinfrastructure for Astronomy VII},  {\em Society of Photo-Optical
  Instrumentation Engineers (SPIE) Conference Series} {\bf 12189}, 1218920
  (2022).

\bibitem{2020SPIE11452E..1TD}
M.~{De Pascale}, A.~{Baruffolo}, B.~{Salasnich}, {\em et~al.}, ``{SHARK-NIR:
  implementation of the instrument control software SHINS},'' in {\em Software
  and Cyberinfrastructure for Astronomy VI},  J.~C. {Guzman} and J.~{Ibsen},
  Eds., {\em Society of Photo-Optical Instrumentation Engineers (SPIE)
  Conference Series} {\bf 11452}, 114521T  (2020).

\bibitem{2018SPIE10707E..1MD}
M.~{De Pascale}, A.~{Baruffolo}, B.~{Salasnich}, {\em et~al.}, ``{Design of
  SHINS: the SHARK-NIR instrument control software},'' in {\em Software and
  Cyberinfrastructure for Astronomy V},  J.~C. {Guzman} and J.~{Ibsen}, Eds.,
  {\em Society of Photo-Optical Instrumentation Engineers (SPIE) Conference
  Series} {\bf 10707}, 107071M  (2018).

\bibitem{2008SPIE.7013E..26H}
T.~M. {Herbst}, R.~{Ragazzoni}, A.~{Eckart}, {\em et~al.}, ``{LINC-NIRVANA: the
  Fizeau interferometer for the Large Binocular Telescope},'' in {\em Optical
  and Infrared Interferometry},  M.~{Sch{\"o}ller}, W.~C. {Danchi}, and
  F.~{Delplancke}, Eds., {\em Society of Photo-Optical Instrumentation
  Engineers (SPIE) Conference Series} {\bf 7013}, 701326  (2008).

\bibitem{2008SPIE.7019E..1TB}
J.~{Berwein}, F.~{Briegel}, W.~{Gaessler}, {\em et~al.}, ``{An SOA developer
  framework for astronomical instrument control software},'' in {\em Advanced
  Software and Control for Astronomy II},  A.~{Bridger} and N.~M. {Radziwill},
  Eds., {\em Society of Photo-Optical Instrumentation Engineers (SPIE)
  Conference Series} {\bf 7019}, 70191T  (2008).

\bibitem{2019A&A...621A...4R}
S.~{Rabien}, R.~{Angel}, L.~{Barl}, {\em et~al.}, ``{ARGOS at the LBT.
  Binocular laser guided ground-layer adaptive optics},'' {\em \aap} {\bf 621},
  A4  (2019).

\bibitem{2012SPIE.8446E..4FL}
J.~M. {Leisenring}, M.~F. {Skrutskie}, P.~M. {Hinz}, {\em et~al.}, ``{On-sky
  operations and performance of LMIRcam at the Large Binocular Telescope},'' in
  {\em Ground-based and Airborne Instrumentation for Astronomy IV},  I.~S.
  {McLean}, S.~K. {Ramsay}, and H.~{Takami}, Eds., {\em Society of
  Photo-Optical Instrumentation Engineers (SPIE) Conference Series} {\bf 8446},
  84464F  (2012).

\bibitem{2018SPIE10705E..16V}
D.~{Vassallo}, J.~{Farinato}, J.~F. {Sauvage}, {\em et~al.}, ``{Validating the
  phase diversity approach for sensing NCPA in SHARK-NIR, the second-generation
  high-contrast imager for the Large Binocular Telescope},'' in {\em Modeling,
  Systems Engineering, and Project Management for Astronomy VIII},  G.~Z.
  {Angeli} and P.~{Dierickx}, Eds., {\em Society of Photo-Optical
  Instrumentation Engineers (SPIE) Conference Series} {\bf 10705}, 1070516
  (2018).

\bibitem{10.5555/932295}
R.~T. Fielding and R.~N. Taylor, {\em Architectural styles and the design of
  network-based software architectures}.
\newblock PhD thesis, University Of California, Irvine  (2000).
\newblock AAI9980887.

\bibitem{2024SPIE13098E..0TR}
D.~{Ricci}, L.~{Cabona}, B.~{Salasnich}, {\em et~al.}, ``{Easy remote
  observations using web interfaces: controlling an Italian telescope from
  Japan, and more},'' in {\em Observatory Operations: Strategies, Processes,
  and Systems X},  C.~R. {Benn}, A.~{Chrysostomou}, and L.~J.
  {Storrie-Lombardi}, Eds., {\em Society of Photo-Optical Instrumentation
  Engineers (SPIE) Conference Series} {\bf 13098}, 130980T  (2024).

\bibitem{2022SPIE12186E..0PR}
D.~{Ricci}, L.~{Cabona}, S.~{Tosi}, {\em et~al.}, ``{Toward the remotization
  and robotization of the OARPAF Telescope},'' in {\em Observatory Operations:
  Strategies, Processes, and Systems IX},  D.~S. {Adler}, R.~L. {Seaman}, and
  C.~R. {Benn}, Eds., {\em Society of Photo-Optical Instrumentation Engineers
  (SPIE) Conference Series} {\bf 12186}, 121860P  (2022).

\bibitem{2000PASP..112.1479R}
D.~{Rouan}, P.~{Riaud}, A.~{Boccaletti}, {\em et~al.}, ``{The Four-Quadrant
  Phase-Mask Coronagraph. I. Principle},'' {\em \pasp} {\bf 112}, 1479--1486
  (2000).

\end{thebibliography}
\bibliographystyle{spiejour}   




\end{document}